\documentclass[trackchanges,twocolumn]{aastex631}

\usepackage{graphicx}
\graphicspath{ {./images/} }
\usepackage{textcomp}
\usepackage{gensymb}
\usepackage{amsmath}
\usepackage{hyperref}
\usepackage{comment}
\usepackage{wrapfig}
\usepackage{upgreek}
\usepackage{longtable}
\usepackage{lipsum}
\usepackage[version=4]{mhchem}

\begin{document}

\title{Extreme Forward Scattering Observed in Disk-Averaged Near-Infrared Phase Curves of Titan}

\author[0009-0008-1520-4272]{Chase Cooper}
\affiliation{Department of Astronomy, University of Arizona, Tucson, AZ 85721, USA}
\affiliation{Lunar and Planetary Laboratory, University of Arizona, Tucson, AZ 85721, USA}
\affiliation{Habitability, Atmospheres, and Biosignatures Laboratory, University of Arizona, Tucson, AZ 85721, USA}

\author[0000-0002-3196-414X]{Tyler D. Robinson}
\affiliation{Lunar and Planetary Laboratory, University of Arizona, Tucson, AZ 85721, USA}
\affiliation{Habitability, Atmospheres, and Biosignatures Laboratory, University of Arizona, Tucson, AZ 85721, USA}
\affiliation{NASA Nexus for Exoplanet System Science Virtual Planetary Laboratory, University of Washington, Box 351580, Seattle, WA 98195, USA}

\author[0000-0002-7755-3530]{Jason W. Barnes}
\affiliation{Department of Physics; University of Idaho; Moscow, ID 83844, USA}

\author[0000-0002-4321-4581]{L.~C. Mayorga}
\affiliation{Johns Hopkins Applied Physics Laboratory, Laurel, MD, 20723, USA}

\author[0009-0004-9747-5623]{Lily Robinthal}
\affiliation{Lunar and Planetary Laboratory, University of Arizona, Tucson, AZ 85721, USA}
\affiliation{Habitability, Atmospheres, and Biosignatures Laboratory, University of Arizona, Tucson, AZ 85721, USA}

\begin{abstract}

Titan, with its thick and hazy atmosphere, is a key world in our solar system for understanding light scattering processes. NASA's \textit{Cassini} mission monitored Titan between 2004 and 2017, where the derived dataset includes a large number of whole disk observations. Once spatially integrated, these whole disk observations reveal Titan's phase-dependent brightness which can serve as an analog for how hazy worlds might appear around other stars. To explore Titan's phase curve, we present a pipeline for whole disk Titan observations acquired by the \textit{Cassini} Visual and Infrared Mapping Spectrometer (VIMS) spanning 0.9--5.1\,$\upmu$m. Application of the pipeline finds over 4,400 quality spatially- and spectrally-resolved datacubes that were then integrated over Titan's disk to yield phase curves spanning 2--165\degree{} in phase angle. Spectra at near-full phase provide a useful approximation for Titan's geometric albedo, thus extending the spectral coverage of previous work. Crescent phase brightness enhancements in the \textit{Cassini}/VIMS phase curves are often more extreme than analogous results seen at optical wavelengths, which can be explained by atmospheric transparency and haze scattering processes. These results provide validation opportunities for exoplanet-focused spectral models and also shed light on how extreme aerosol forward scattering could influence exoplanet observations and interpretations.

\end{abstract}


\section{Introduction} \label{sec:intro}

Observations of exoplanets have revealed an impressive diversity in planetary atmospheres. The atmospheres of many Jupiter- to Neptune-sized exoplanets have been successfully studied by looking at features in transit \citep[e.g.,][]{Madhusudhan_2020,Kesseli_2022,Zhang_2022,Madhusudhan_2023,Rustamkulov_2023}, thermal emission \citep{Charbonneau_2005,Knutson_2006,Kreidberg_2014,Stevenson_2014}, and/or reflected light \citep{Esteves_2015,Hoeijmakers_2018,Lendl_2020,Hooton_2022,Brandeker_2022} spectra. For a subset of worlds\,---\,especially hot Jupiters\,---\,observations of their changing thermal brightness with star-planet-observer (i.e., phase) angle has helped to constrain key aspects of atmospheric thermal structure and circulation \citep[for a convenient summary of targets, see Table~1 in][]{Parmentier_2018}.

In reflected-light direct imaging, current telescopes are generally unable to resolve planetary companions and study phase curves, especially for rocky exoplanets \citep[see, e.g.,][]{Wang_2017}. In the future, though, NASA's under-development Habitable Worlds Observatory \citep[HWO; ][]{feinbergetal2024} will provide the high-contrast imaging capabilities required to study a wide range of planet types, including clement terrestrial worlds. Critically, repeat visits to planetary systems and targets will enable HWO to build up phase curves for many worlds over the duration of the mission.

Studying reflected-light phase curves can constrain surface and/or atmospheric properties of a planet. In general, an object's brightness will decrease as phase angle increases, as progressively less of the planetary disk is illuminated from the observer's perspective. Moving beyond simple illumination effects, the shape of phase curves can be affected by phenomena that can reveal details of the atmospheric and/or surface state. For example, phase curves of Venus contain a feature at high phase angles due to sulfuric acid droplets in the atmosphere \citep{Arking_1968}, while optical phase curves of Titan show that the moon's haze is responsible for forward scattering of optical light \citep{GarciaMunoz_2017}. Phase curves of solar system objects can further serve as testing grounds for future direct observations of exoplanets and the production of their phase curves, as has been done with both Jupiter \citep{Mayorga_2016,Heng_2021} and Saturn \citep{Hedman_2015,Dyudina_2016}.

Modeling focused studies of exoplanet reflected-light phase curves have emphasized the potential for detecting specular reflection from liquid water oceans (i.e., glint), thus revealing a habitable surface environment \citep{Williams_2008}. \citet{Robinson_2010} used Earth models to create phase curves of Earth both with and without glint contributions, and found that Earth with glint can appear twice as bright as Earth without glint at crescent phase. Detecting glint effects in nearby exoplanet phase curves could be within the capabilities of HWO \citep{Vaughn_2023}. The presence of surface oceans may also be inferred from measuring reflected light polarization as a function of phase, indicated by a peak polarization just past half-phase \citep{Zugger_2010}.

Most fundamentally, an ocean glint signature in a planetary phase curve is revealed through forward scattering. Thus, the previously mentioned haze forward scattering detections from \citet{GarciaMunoz_2017} could present a potential false positive for glint. This connection between glint and Titan is particularly interesting given the collection of studies that use glint from Titan to study its seas \citep{Stephan_2010,Barnes_2011,Soderblom_2012,Barnes_2013,barnes_2014}. The work presented here further explores aerosol forward scattering in Titan phase curves.

In the solar system, Titan stands out among planetary objects as an example of a rocky/icy body with a thick atmosphere. Due to the large semi-major axis of the Saturn system in comparison to that of the Earth, both ground and Earth-orbiting telescopes can only view Titan at phase angles below about $6.5\degree$. Fortunately, access to viewing angles not obtainable from Earth was provided by NASA's \textit{Cassini} spacecraft, which made orbital observations of Saturn, its rings, and its moons from 2004 to 2017 \citep{CassiniTitan_2019}. \textit{Cassini} made regular flybys of Titan, taking data of the moon in the ultraviolet, visible, infrared, and radio wavelength regimes. Among the instruments on board the probe was the Visual and Infrared Mapping Spectrometer (VIMS) instrument, which took spatially resolved images of Titan in the optical (0.3--0.9\,$\upmu$m) and near-infrared (0.9--5.1\,$\upmu$m) wavelength regimes \citep{Brown_2004}.

The primary focus of this work is to use \textit{Cassini}/VIMS observations of Titan to generate phase curves that span the near-infrared wavelength regime, thereby complementing the optical studies of \citet{GarciaMunoz_2017}. Section \ref{sec:methods} covers data acquisition and analysis using available resources, as well as the pipeline we produced to reduce spatially resolved images to disk-averaged reflectivity measures. In Section \ref{sec:results}, near-infrared reflected-light phase curves of Titan are presented and key features are noted. Section \ref{sec:discussion} explores possible explanations of the two phenomena mentioned in Section \ref{sec:results}, and compares our findings to those of \citet{GarciaMunoz_2017}. Finally, Section \ref{sec:conclusion} will summarize the findings of this work, as well as connect these findings to ongoing research of exoplanets.

\section{Methods} \label{sec:methods}

Section \ref{subsec:vet} details the process of acquiring and filtering VIMS data. Section \ref{subsec:disk} describes our automated disk detection process. Section \ref{subsec:if} gives an overview of how disk-averaged measurements are derived from VIMS pixel data.

\subsection{File Vetting and Calibration} \label{subsec:vet}
\begin{deluxetable}{cc}[b]
    \tablecaption{Filters used to remove unusable data \label{table:filter}}
    \tablenum{1}
    \tablehead{\colhead{Filter Criteria} & \colhead{\# of cubes removed}}
    \startdata
    Empty file &  35 \\
    Likely swaths &  45,333 \\
    Missing band data &  97 \\
    Failed edge detection &  4,104 \\
    Failed calibration &  83 \\
    Titan disk exceeds FOV &  6,643 \\
    Incomplete disk &  1,217 \\
    Manual sorting & 372\\
    \hline
    Total invalid cubes &  57,884 \\
    Valid cubes &  4,492 \\
    \enddata
\end{deluxetable}

Data were acquired using the PDS Image Atlas on NASA's Planetary Data System. In the rest of this paper, downloaded files are referred to as ``cubes.'' Cubes contain general information about the time and duration of the observation, as well as the location and orientation of \textit{Cassini} at time of acquisition. Each cube includes 352 two-dimensional arrays---one array for each VIMS band---and each array contains spatially-resolved counts accumulated during the cube's exposure time. Arrays 1--96 correspond to VIMS bands in optical light, while arrays 97--352 correspond to bands at infrared (IR) wavelengths \citep{Brown_2004}. Our analyzed cubes were acquired throughout the full duration of the \textit{Cassini} mission (2004--2017), were taken at phase angles from $2\degree$ to $165\degree$, and contained imaging data at all infrared bands covered by the VIMS-IR camera (0.9--5.1\,$\upmu$m). A total of 62,376 cubes of Titan were downloaded.

We placed a constraint that cubes be at least 12 pixels to a side in size. Critically, this criterion removed the many swaths in our dataset\,---\,long, narrow scans of Titan's surface taken during close approaches. We also exclude cubes taken at distances where the apparent diameter of Titan exceeds the maximum cube size of 64 pixels to a side. The diameter of Titan in pixels is calculated as:
$$
2R_{\text{pixel}}=\frac{2}{S}\tan(\frac{R}{d})
$$
where $S=5\cdot10^{-4}$ radians is the field of view of a VIMS pixel on a side, $R$ is the radius of Titan in km, and $d$ is the distance of the spacecraft at time of observation. By letting $2R_{\text{pixel}}=64$ and solving for $d$, we get a lower distance limit of approximately 180,000 km. Cubes missing data from more than 10\% of VIMS bands were also removed. 

We then used a simple Canny edge detection method to test for structures in cube data, and excluded cubes showing no structure. Canny edge detection algorithms are a family of algorithms that identify edges and boundaries in pixelated images \citep{Canny_1986}. In the case of Titan images, Canny edge detection returns an array that identifies pixels which either lie on the edge of the disk of Titan or on the boundaries between geographical regions. By requiring that cubes have structure as determined by this algorithm, we exclude cubes that do not contain the disk of Titan, that zoom in on a subsection of the disk, or have erroneous data. Edge detection was also used to remove cubes not containing the full disk of Titan. Finally, remaining cubes were manually inspected to ensure they met our criteria. Table~\ref{table:filter} breaks down the different criteria used to sort the cubes, in the order they were applied, and the amount of cubes removed by each criterion.  Once the filtering process was finished, our dataset contained 4,492 cubes.

All cubes were passed through the multi-step calibration process described by \citet{LeMoulic_2019} using the USGS ISIS3 software package \citep{isis}. The calibration process formats each cube to be ISIS3-compliant, performs background subtraction and noise filtering, removes sources of error such as imaging artifacts and cosmic rays, and adds SPICE data. SPICE data includes information on the location and orientation of the spacecraft during its mission, as well as instrument details, target information and more \citep{ACTON199665,ACTON20189}. The calibration also converted data numbers, which are correlated with photon counts, to spatially-resolved $I/F$ values, a unit of measurement of an object's reflectivity per steradian (see Section~\ref{subsec:if}). Cubes were calibrated this way in order to be compatible with the \texttt{pyvims} Python library, the primary tool used for the analysis of cube data. The \texttt{pyvims} software suite \citep[described in ][]{LeMoulic_2019} facilitates the analysis of VIMS cubes and can extract from a calibrated cube the full suite of ephemeris and pointing quantities, including sub-spacecraft coordinates and phase angle.

VIMS data taken in the 0.9--1.3\,$\upmu$m wavelength range suffered from extensive saturation issues, especially at high phase angles. The cause is likely longer integration times \---\ cubes with longer exposure times showed a greater degree of saturation. The phase curves produced from these data show greater spread and do not closely resemble a continuous curve. As a result, we are unable to properly analyze most of these data. We include these data in later figures for comparisons with previous works. Similar saturation issues extended to a few cubes at longer wavelengths at higher phase angles, and these cubes were excluded from our dataset via manual sorting.

\subsection{Automated Disk Detection}\label{subsec:disk}

\begin{figure*}[t]
    \centering
    \plotone{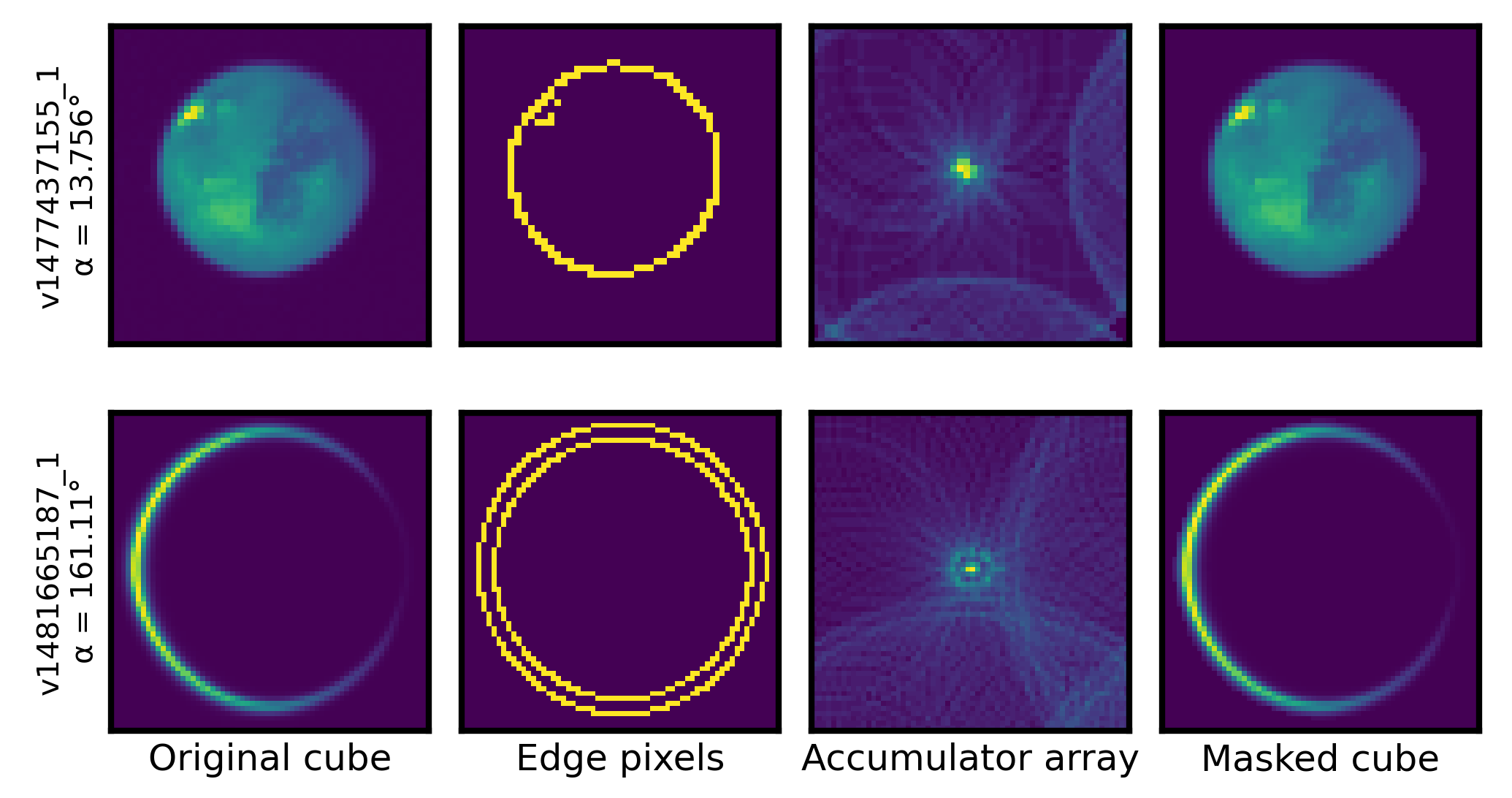}
    \caption{Major steps in the disk detection process on two cubes: (far left) an unaltered cube; (center left) edge pixels identified using Canny edge detection \---\ some pixels along terrain boundaries are falsely identified as edge pixels; (center right) the accumulator array after applying a circle Hough transform; (far right) the final array, containing only data from the disk of Titan. Data were taken at 2\,$\upmu$m.}
    \label{fig:diskdetection}
\end{figure*}

The automatic detection of Titan within images was achieved using a pipeline written in Python, and mirrors the disk detection process used by \citet{Strauss_2024}. A visual summary of the process can be seen in Figure~\ref{fig:diskdetection}. First, the radius of Titan $R_{\text{Titan}}$ is determined by adding the wavelength-dependent atmospheric height of Titan \citep{Robinson_2014} to the solid body radius of $2,575$\,km \citep{Zebker_2009}. Effective height data at wavelengths corresponding to VIMS-IR bands 345--352 were absent, so the effective height used for band 344 was applied to these bands as well. Then Titan's radius in units of VIMS pixels is calculated as $R_{\text{pxl}}=\lceil{R_{\text{Titan}}/S}\rceil$, where $S$ is the average surface resolution of the image in kilometers per pixel as determined with \texttt{pyvims}. This radius is increased by 1 in cases where the phase angle of the image is above $120\degree$, because the atmosphere appears much brighter due to forward scattering of light by atmospheric aerosols, which increases the apparent radius of Titan. Increasing this radius by more than 1 pixel did not substantially change the disk-averaged measurements.

% Next, Canny edge detection \citep{Canny_1986} is applied to the image. Canny edge detection algorithms are a family of algorithms that identify edges and boundaries in pixelated images. In the case of images, Canny edge detection returns an array that identifies pixels which likely lie on the edge of the disk of Titan.

Next, we used a Canny edge detection algorithm to identify pixels on the edge of Titan's visible disk. A circle Hough transform is then used to locate the (approximate) center of the disk \citep{xie_2002}. In the circle Hough transform, an accumulator array with the same dimensions as the image is created. Circles with a radius of $R_{\rm pxl}$ pixels are overlaid on the array, centered on each edge pixel location in the edge pixel array, with each pixel covered by a circle having its accumulator array value incremented. After each circle has been overlaid, the pixel with the highest value is taken to be the center of the disk, and an array is created where each pixel within $R_{\text{pxl}}$ pixels of the center pixel identified above has a value of 1, and all other pixels have a value of 0. By multiplying the original image array by this masking array, the resulting array only contains data from pixels lying on the disk of Titan.

\subsection{$A_{\rm g}\Phi(\alpha)$ Calculation} \label{subsec:if}

The quantity $A_{\rm g}\Phi(\alpha)$ is the product of an object's geometric albedo, $A_g$, and its planetary phase function, $\Phi(\alpha)$. In planetary science, $A_{\rm g}\Phi(\alpha)$ is often used as a metric of a planetary object's reflectivity, as it quantifies the ratio of the radiance received from an object to the irradiance that object receives. As a reflectance, it is a unitless quantity. The cube calibration process described above converts data numbers from the raw \textit{Cassini} data to $I/F$ values, with units of inverse steradians. Because we aim to simulate point-like observations of Titan rather than the spatially resolved observations provided by \textit{Cassini}, we convert the spatial $I/F$ values across Titan's disk to a single ``disk averaged'' quantity. The disk-averaged $A_{\rm g}\Phi(\alpha)$ for Titan is calculated as,

\begin{equation}
    A_{\rm g}\Phi(\alpha) = \pi \frac{\bar{I}}{F} = \frac{d_{\rm sc}^2\Omega}{\pi R^2} \sum_i (I/F)_{i}
\end{equation}
where $\bar{I}/F$ is the disk-averaged intensity-to-incident flux ratio, $d_{\rm sc}$ is the distance from \textit{Cassini} to the center of Titan in \,km, $\Omega=2.5 \times 10^{-7}$ is the solid angle of a VIMS pixel in steradians, $R$ is the solid body radius of Titan in \,km, and the sum is over individual $I/F$ values of all pixels on the disk. \texttt{pyvims} does not provide information on errors in individual pixel measurements, however integrating over the disk significantly reduces random errors.
% The quantity $I/F$ is equivalent to geometric albedo times planetary phase function, often written as $A_g\Phi(\alpha)$ in the literature.

\section{Results} \label{sec:results}

\begin{figure*}
    \centering
    \includegraphics[width=\linewidth]{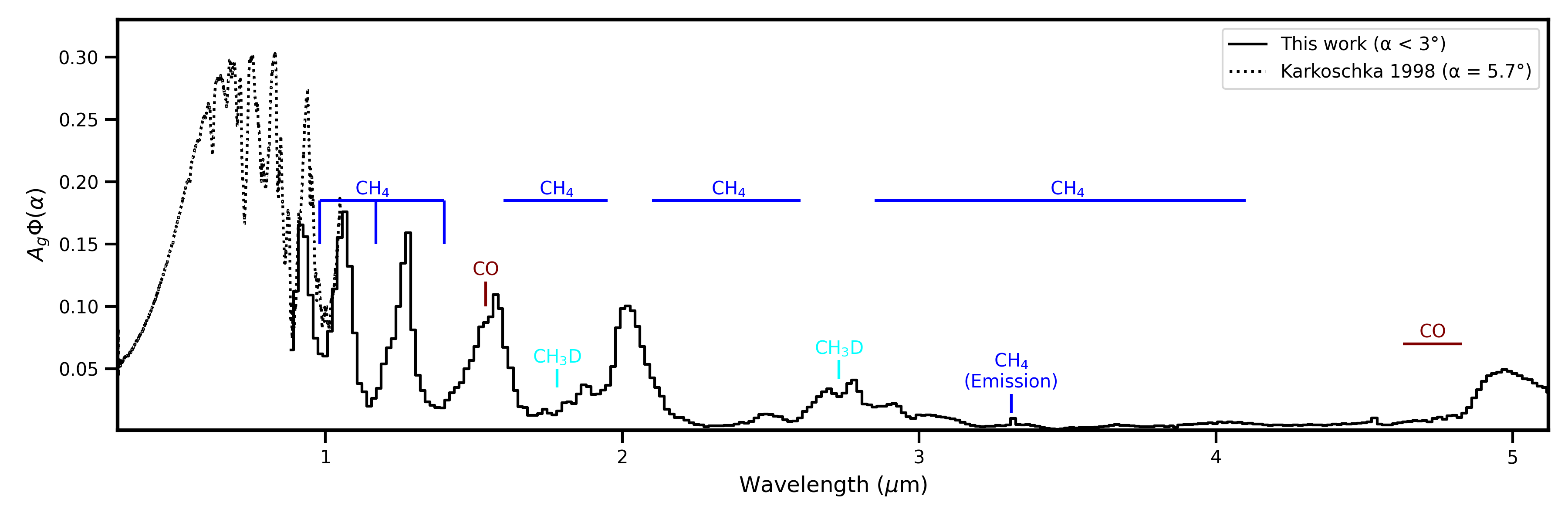}
    \caption{A near-full phase $A_{\rm g}\Phi(\alpha)$  spectrum of Titan. Data from \citet{Karkoschka_1998} are included for comparison. The spectrum was obtained by taking the disk-averaged spectra of 10 cubes with phase angle $\alpha<3\degree$. Data were acquired at very low phase angle and, thus, approximate the geometric albedo spectrum of Titan. Certain spectral features and their sources are also identified. Features at 3.31\,$\upmu$m and 4.54\,$\upmu$m are discussed in Section \ref{subsec:disc_4.54}. Features in the \textit{Cassini} data are consistent with the model results from \citet{Es-sayeh_2023} and recent \textit{James Webb Space Telescope} data from \citet{Nixon_2025}.}
    \label{fig:labelled}
\end{figure*}

\begin{figure}
    \includegraphics[width=\linewidth]{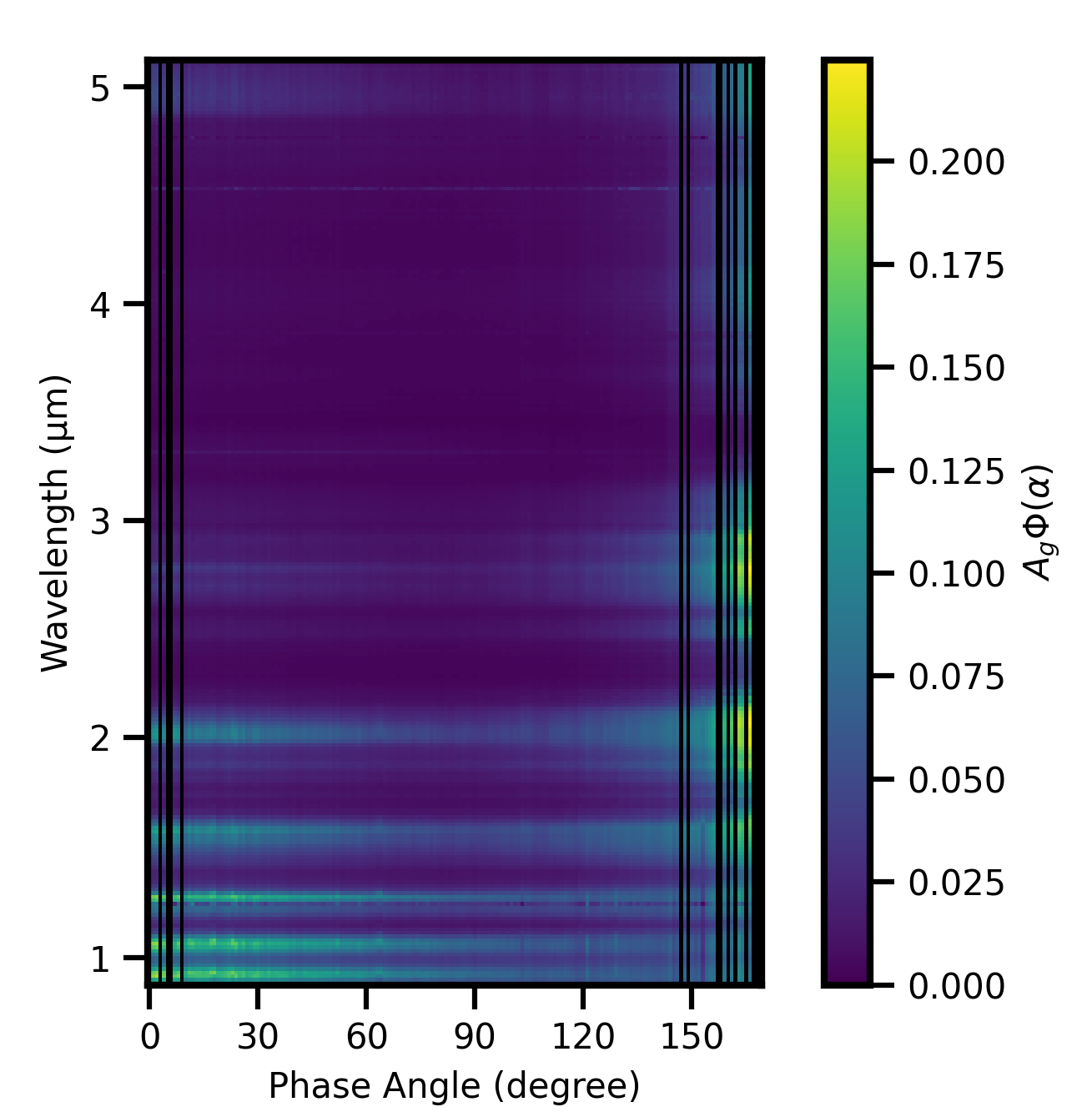}
    \caption{$A_{\rm g}\Phi(\alpha)$ of Titan as a function of wavelength and phase angle. Vertical black lines indicate phase angles at which there were no cubes to consider. Bright horizontal bands are associated with continuum between strong methane absorption bands.}
    \label{fig:heatmap}
\end{figure}

A disk-averaged $A_{\rm g}\Phi(\alpha)$ spectrum recorded at the smallest phase angle in our dataset (2\degree) is shown in Figure~\ref{fig:labelled}. As the variation of $A_{\rm g}\Phi(\alpha)$ with phase at these low phase angles is small, this spectrum is a reasonable stand-in for a geometric albedo spectrum (which is formally defined at full phase). Also included for comparison, and extension, are optical data from \citet{Karkoschka_1998}, which corroborate the full-phase values derived by \citet{GarciaMunoz_2017}. The discrepancy between the findings of this work and that of Karkoschka is explained by the saturation issues mentioned at the end of Section~\ref{subsec:vet}. Figure~\ref{fig:labelled} highlights just the VIMS (near) geometric albedo spectrum with notable bands due to \ce{CH4}, \ce{CO}, and \ce{CH3D} indicated. Continuum regions between absorption bands show a generally decreasing $A_{\rm g}\Phi(\alpha)$ with wavelength, consistent with the trend of decreasing haze single scattering albedo with increasing wavelength \citep{Tomasko_2008}. Notable atmospheric windows with sensitivity to the deeper atmosphere/surface occur at 0.94, 1.08, 1.28, 1.6, 2.0, 2.7, 2.8, and 5.0\,$\upmu$m.

% \begin{figure*}[t]
%     \centering
%     \includegraphics[width=\linewidth]{images/albedogeo.png}
%     \caption{A near-full phase $A_g\Phi(\alpha)$  spectrum of Titan. Data from \citet{Karkoschka_1998} are included for comparison. The spectrum was obtained by taking the disk-averaged spectra of 10 cubes with phase angle $\alpha<3\degree$. Data were acquired at very low phase angle and, thus, approximate the geometric albedo spectrum of Titan. Features at 3.31\,$\upmu$m and 4.54\,$\upmu$m are discussed in Section \ref{subsec:disc_4.54}. 
%     }
%     \label{fig:geomalb}
% \end{figure*}

Figure \ref{fig:heatmap} shows the value of $A_{\rm g}\Phi(\alpha)$ for Titan as a function of wavelength and phase angle. These disk-averaged $A_{\rm g}\Phi(\alpha)$ values are not normalized to full phase, so are not formal phase functions. Cubes were binned by phase angle into one-degree bins, and for bins with multiple cubes an average $A_{\rm g}\Phi(\alpha)$ value was computed. Vertical black stripes represent phase angle bins with no viable cubes. The region on the right edge of the figure indicates where significant forward scattering causes Titan to appear bright, as discussed above. These regions largely correspond to continuum outside of Titan's atmospheric absorption features.

\begin{figure*}
    \centering
    \includegraphics[scale=0.60]{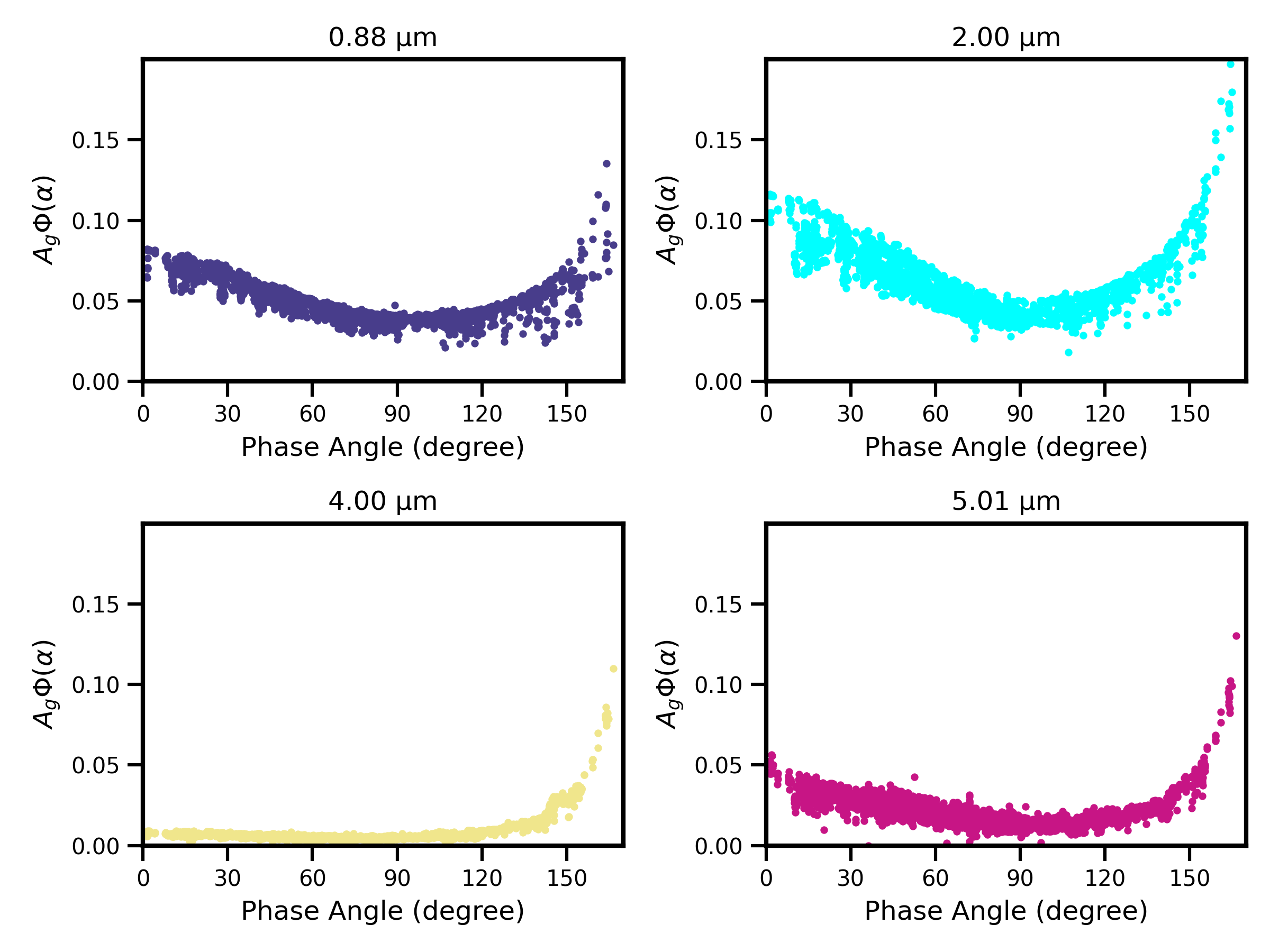}
    \caption{A suite of Titan phase curves taken at different wavelengths. At each wavelength, extreme forward scattering is apparent at high phase angles, even at wavelengths where the disk of Titan appears dim when observing at low phase angles (e.g. 4.0\,$\upmu$m). The effect is strong enough that Titan's disk-averaged brightness is greater at high phase than at low phase for almost all wavelengths. The top left phase curve, taken at 0.88\,$\upmu$m, suffers from saturation-related issues at high phase.}
    \label{fig:quadphasecurves}
\end{figure*}

Figure~\ref{fig:quadphasecurves} shows a selection of phase curves at select wavelengths produced by our pipeline. The phase curve of Titan is decidedly not Lambertian, in contrast to those of other major Saturnian satellites which are better-approximated by a Lambert phase function \citep{Buratti_1984}. At every wavelength, the brightness of Titan follows a smooth curve that initially decreases as the phase angle increases from roughly $10\degree$ to $100\degree$. At phase angles above about $100\degree$, the brightness of the disk then increases due to forward scattering of sunlight by the atmosphere. At phase angles above about $140\degree$, Titan appears brighter than when observed at near-full phase. While this effect has been observed in the 0.3--1\,$\upmu$m range before \citep{Tomasko_2008,Doose_2016,GarciaMunoz_2017}, our analysis confirms that significant forward scattering occurs at most infrared VIMS bands. The larger spread in $A_{\rm g}\Phi(\alpha)$ at smaller phase angles in the 2\,$\upmu$m plot in Figure~\ref{fig:quadphasecurves} is apparent in all phase curves recorded at wavelengths with sensitivity to the deep atmosphere and/or surface and will be discussed later.

\begin{figure*}
    \centering
    \includegraphics[width=\linewidth]{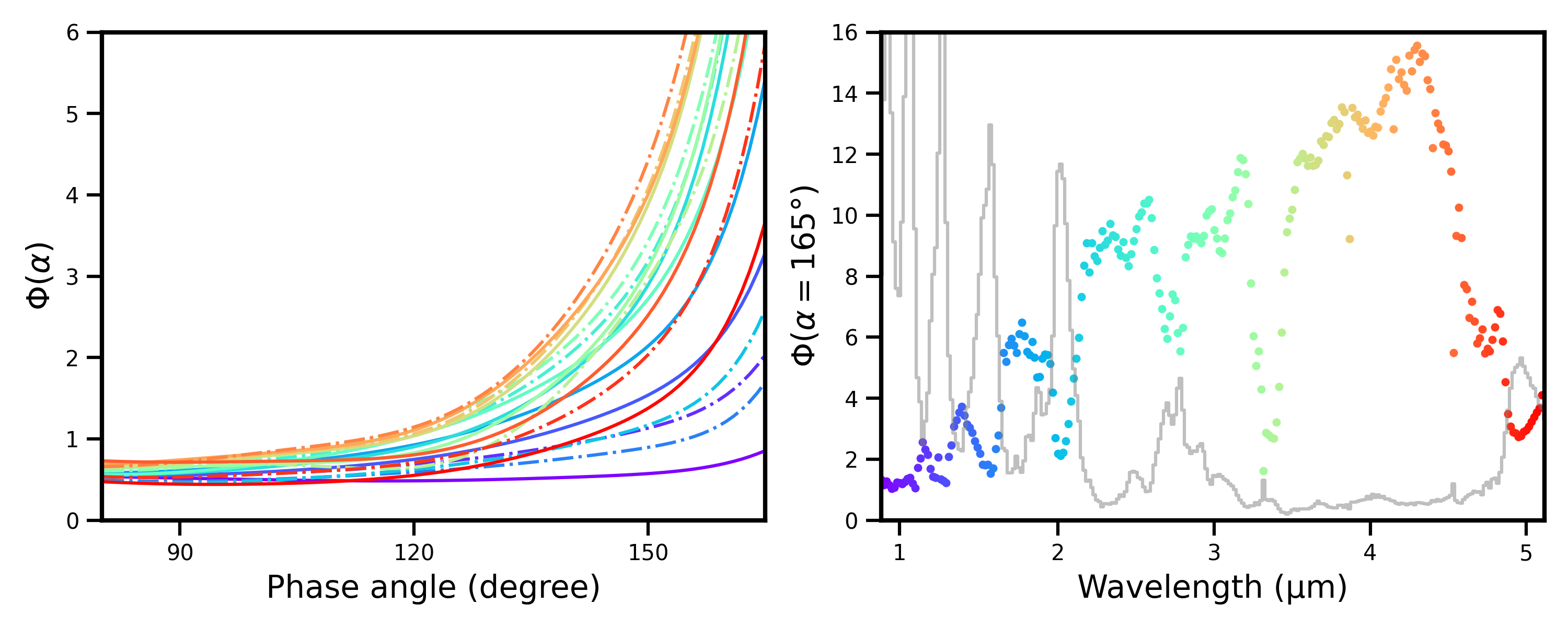}
    \caption{Left: A selection of approximate planetary phase functions from across the near-infrared spectrum accessible to VIMS-IR. The color of fits are indicative of the VIMS-IR band at which they were generated. Curves are normalized to 10$\degree$ due to a sparse set of data at phase angles below 10$\degree$. Right: The value of Titan's planetary phase function $\Phi(\alpha)$ at the max phase angle in our dataset, $\alpha=165\degree$. The magnitude of this value broadly scales with wavelength. Some dips within the structure correspond to spectral continuum regions, similar to features in the infrared spectrum of Titan (grey; spectrum averaged over all cubes)}
    \label{fig:curves_peaks}
\end{figure*}

Polynomial fits of degree 10 to select phase curves at wavelengths between 0.88$\upmu$m and 5.12$\upmu$m are shown in Figure~\ref{fig:curves_peaks}, normalized to lowest-phase $A_{\rm g}\Phi(\alpha)$ values, thereby approximating the planetary phase function $\Phi(\alpha)$. All phase curves show a gradual decrease from roughly 2$\degree$--80$\degree$ before leveling out. Significant brightness surges occur beginning around 130\degree. The effect is weakest at shorter wavelengths, though for some bands this muting may be due to aforementioned saturation issues. Some normalized phase curves taken at wavelengths between 3 and 4$\upmu$m show nearly an order of magnitude increase between near-full and crescent phases as a result of significant forward scattering by atmospheric aerosols.

\begin{figure}
    \centering
    \includegraphics[width=\linewidth]{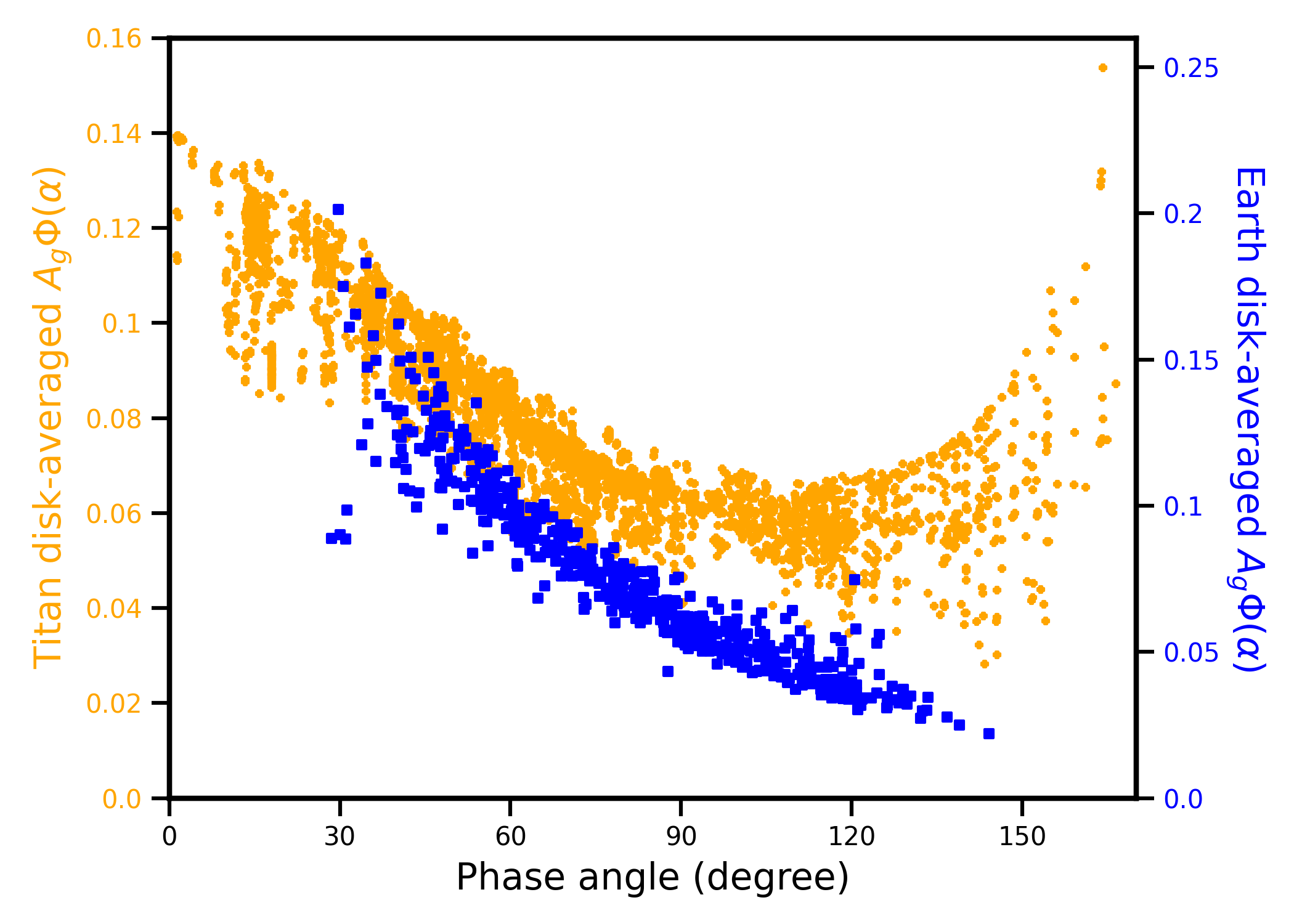}
    \caption{Comparison of a Titan phase curve at a continuum wavelength (0.9\,$\upmu$m) to Earth's broadband visible phase curve \citep[data from ][]{Qiu_2003,Palle_2003}.}
    \label{fig:earthcomp}
\end{figure}

\begin{figure}
    \centering
    \includegraphics[width=\linewidth]{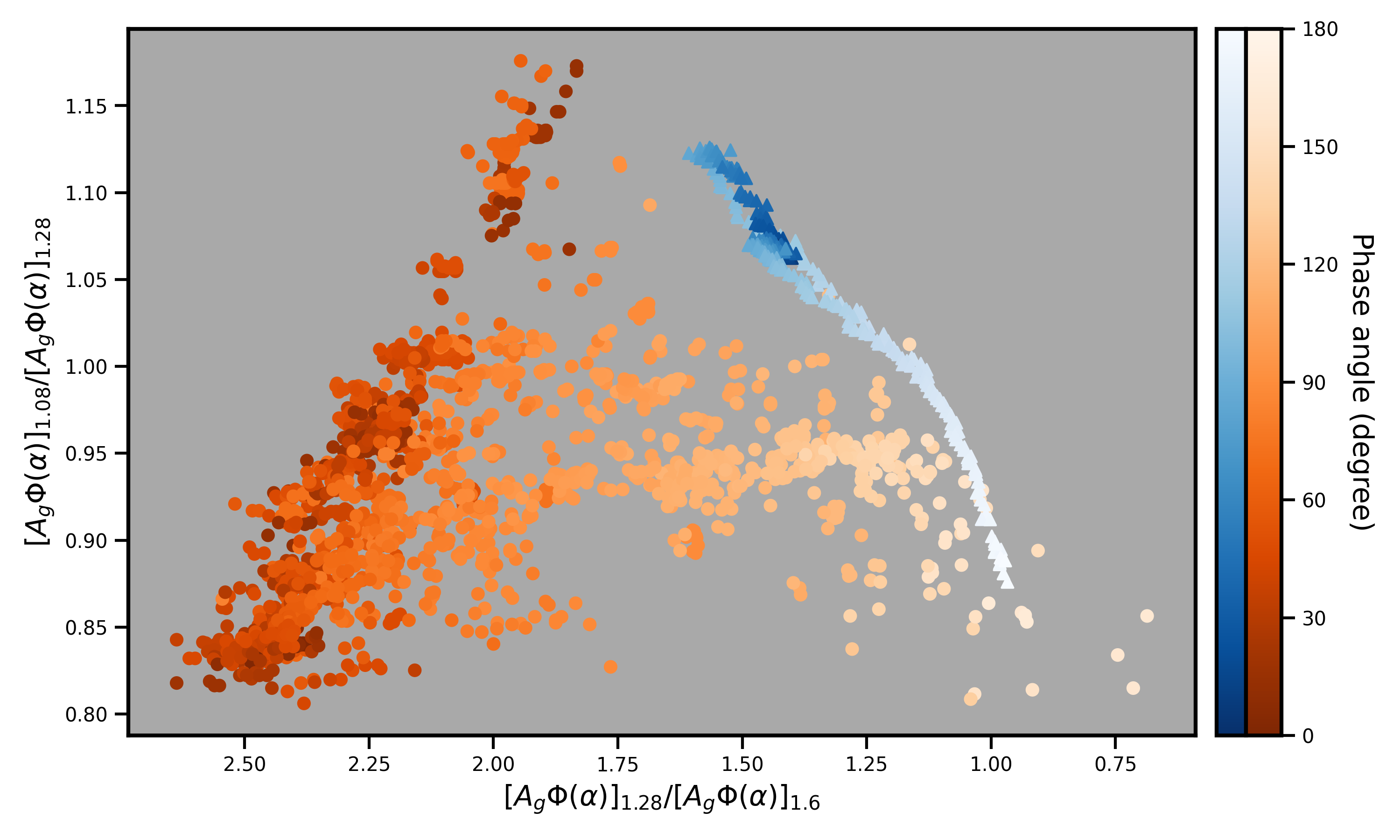}
    \caption{Comparison of Titan (orange) and Earth (blue) phase-dependent brightness in color-color space. Earth values are from spectral models in \citet{Robinson_2010}. Adopted spectral elements at 1.08\,$\upmu$m, 1.28\,$\upmu$m, and 1.60\,$\upmu$m are continuum for both worlds and within the anticipated spectral coverage for HWO. Point saturation indicates phase angle, indicating that Earth and Titan separate well in this color-color space except at very large phase angles.}
    \label{fig:colcol}
\end{figure}

Figures~\ref{fig:earthcomp} and \ref{fig:colcol} emphasize the value of disk-averaged Titan observations as an analog for a hazy exoplanet via comparisons to an Earth phase curve and Earth color-color data, respectively. Published measurements of Earth's phase curve at wavelengths corresponding to VIMS do not exist, so the broadband visible (0.4--0.7\,$\upmu$m) Earthshine data are shown \citep{Qiu_2003,Palle_2003}. Similarly, phase-dependent color measurements for Earth at VIMS-analogous wavelengths do not exist, so we adopt high-fidelity, phase-dependent simulations from the Virtual Planetary Laboratory 3-D Spectral Earth Model \citep{Robinson_2010}.

\section{Discussion} \label{sec:discussion}

\subsection{Phase Curve Structure and Comparisons to Optical Results} \label{subsec:disc_comp}

At all wavelengths, the phase curves presented here share the same general structure. Near full-phase, disk-averaged $A_{\rm g}\Phi(\alpha)$ starts modest and decreases as phase angle increases, with minimum disk-averaged $A_{\rm g}\Phi(\alpha)$ measurements occurring near $\sim100$\degree. Curves then sharply increase, beginning around $\sim$140\degree{} phase. Maximum disk-averaged $A_{\rm g}\Phi(\alpha)$ measurements occur at high phase at nearly all wavelengths; at the shortest wavelengths accessible to VIMS, disk-averaged $A_{\rm g}\Phi(\alpha)$ at lowest and highest phase are comparable, though this is partially due to aforementioned saturation problems at these wavelengths. At atmospheric window wavelengths (e.g., the 2\,$\upmu$m plot in Figure~\ref{fig:quadphasecurves}), photons are scattered fewer times overall, in many cases only once, and the back-scattering peak in the haze's scattering phase function \citep{WEST1991330} results in stronger brightening between quadrature phase and full phase.

The wavelength dependence of the normalized crescent phase peak $A_{\rm g}\Phi(\alpha)$ is primarily controlled by aerosol scattering and gas absorption optical depths. Previous estimates on the size of Titan aerosols are on the order of 1\,$\upmu$m \citep{Rages_1983,Waite_2007,Tomasko_2008}, and strong forward-scattering is to be expected when particle sizes are approximately the same size as the wavelength of incident light. However, a complete description of the scattering behavior of Titan's hazes requires consideration of the fractal aggregate nature of these hazes \citep[see][]{WEST1991330,Tomasko_2008, LAVVAS2010832}. The strong forward peak in the aerosol single-scattering phase function explains the general behavior of larger disk-averaged $A_{\rm g}\Phi(\alpha)$ values at crescent phases versus near full phase given the ubiquity of haze aerosols in Titan's atmosphere \citep{GarciaMunoz_2017}. Increasing atmospheric aerosol transparency at longer wavelengths causes scattering to tend towards the single-scattering regime, which explains the general trend with wavelength in Figure~\ref{fig:curves_peaks}. The figure breaks from this trend around 4.3\,$\upmu$m, though the exact reason for this is unknown. At wavelengths with strong gas absorption, photons incident at the near-full phase geometry would typically require multiple scatterings to escape the atmosphere so are, instead, absorbed along such a path. At crescent phases, though, only a limited number of scattering events are required to direct photons towards the observer (spacecraft), making observations at these wavelengths and phase less sensitive to gas absorption in the deeper atmosphere and more sensitive to single-scattered radiation. A comprehensive spectral model of aerosol scattering could further explain these results but falls beyond the scope of this work.

\begin{figure}
    \centering
    \includegraphics[width=0.85\linewidth]{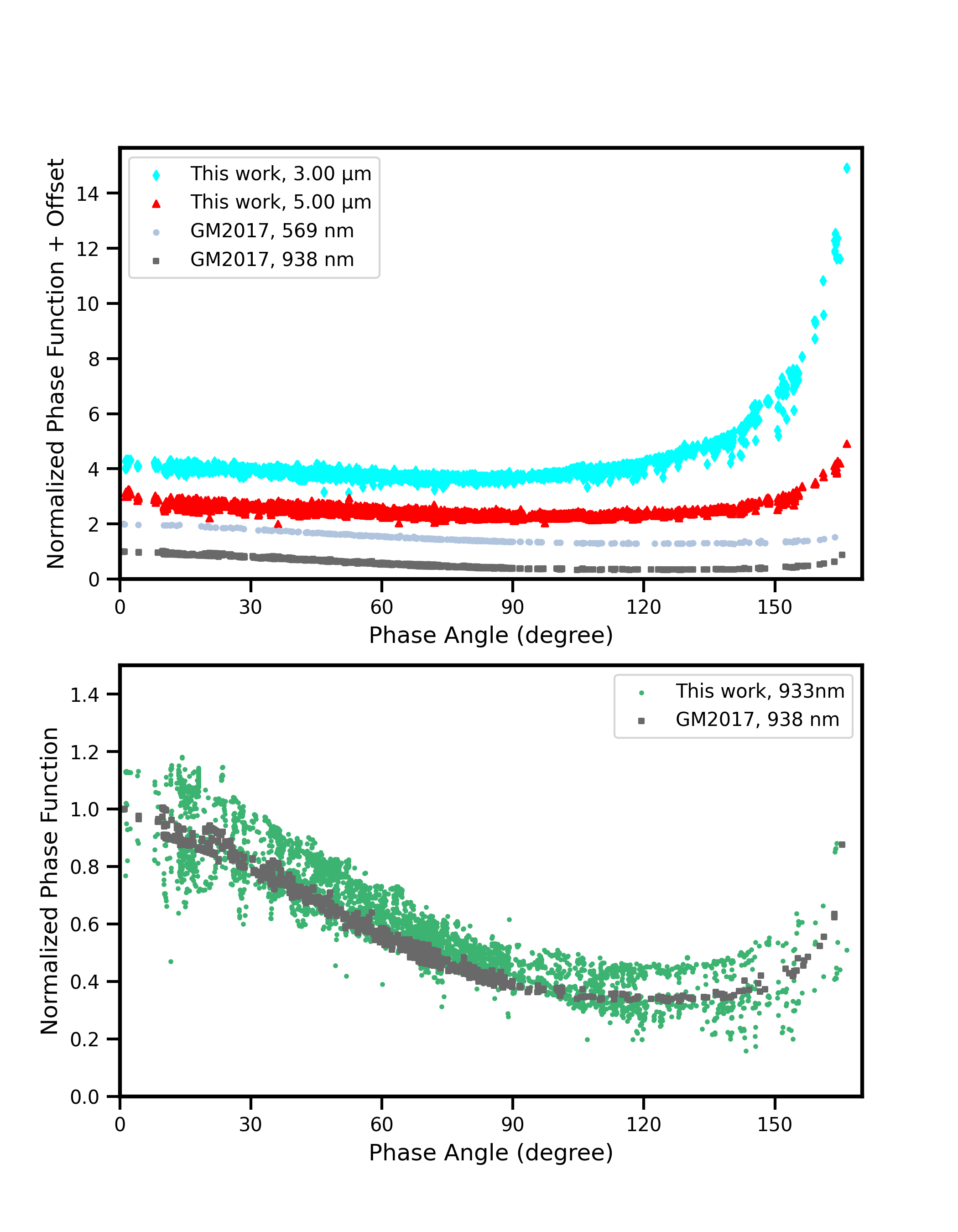}
    \caption{Top: A comparison of two phase curves from the present work (3.00\,$\upmu$m in teal; 5.00\,$\upmu$m in red) and two phase curves from \citet{GarciaMunoz_2017} (569\,nm in light gray; 938\,nm in dark gray). Phase curves are normalized to their respective lowest-phase observation and offset from one another for clarity. Bottom: A comparison of the 938\,nm phase curve from \citet{GarciaMunoz_2017} with our phase curve at 933\,nm. The curves share a similar shape, though our phase curve has much more noise because of saturation issues.}
    \label{fig:GMcomp}
\end{figure}

The general shape of near-infrared phase curves produced with our pipeline matches that of the optical phase curves produced by \citet{GarciaMunoz_2017}. A comparison of our results to data from this earlier work is shown in Figure~\ref{fig:GMcomp}. Both sets demonstrate intense forward scattering from Titan's atmospheric aerosols. The onset of brightness surges occur at similar phase angles in our phase curves ($\sim$135\degree) as in those of \citet{GarciaMunoz_2017} ($\sim$150\degree). However, the relative strength of forward scattering is evidently much different between the two wavelength regimes. Several near-infrared phase curve fits across our sampled phase angle range (2$\degree$--165$\degree$) show approximate $\Phi(\alpha)$ values of 10 or more at high phase; optical light phase curve values across the same intervals are of order unity (consistent with results at the shortest VIMS-IR wavelengths). The differences in the extent of the crescent phase peak in Titan's phase curves at optical versus near-infrared wavelengths aligns with the earlier physical explanation rooted in haze scattering.

\subsection{Notable Spectral Features}\label{subsec:disc_4.54}

\begin{figure}
    \centering
    \includegraphics[width=\linewidth]{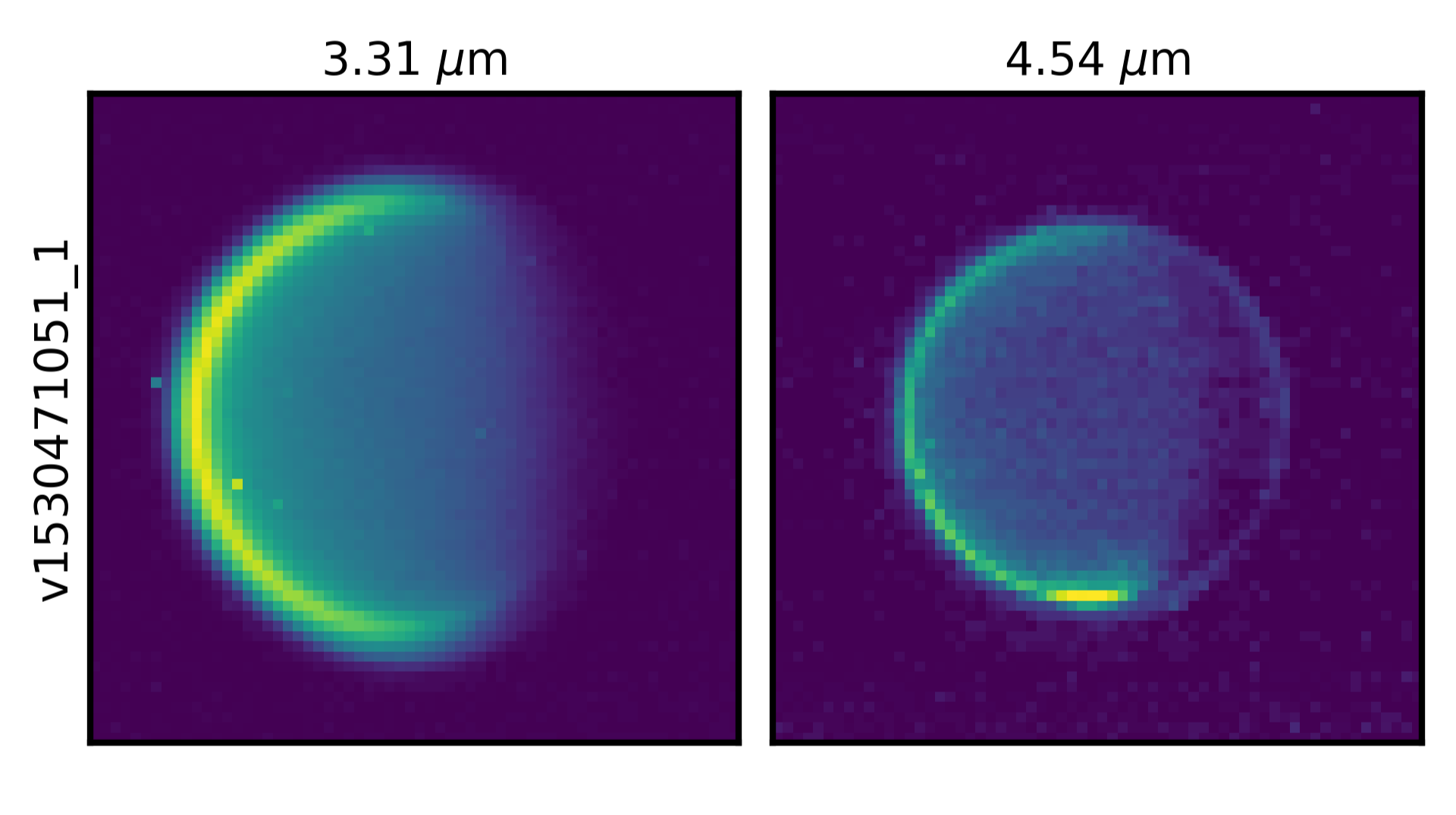}
    \caption{Images of Titan at 3.31\,$\upmu$m (left) and 4.54\,$\upmu$m (right), showing source locations of emission.}
    \label{fig:emission_img}
\end{figure}

\begin{figure}
    \includegraphics[width=\linewidth]{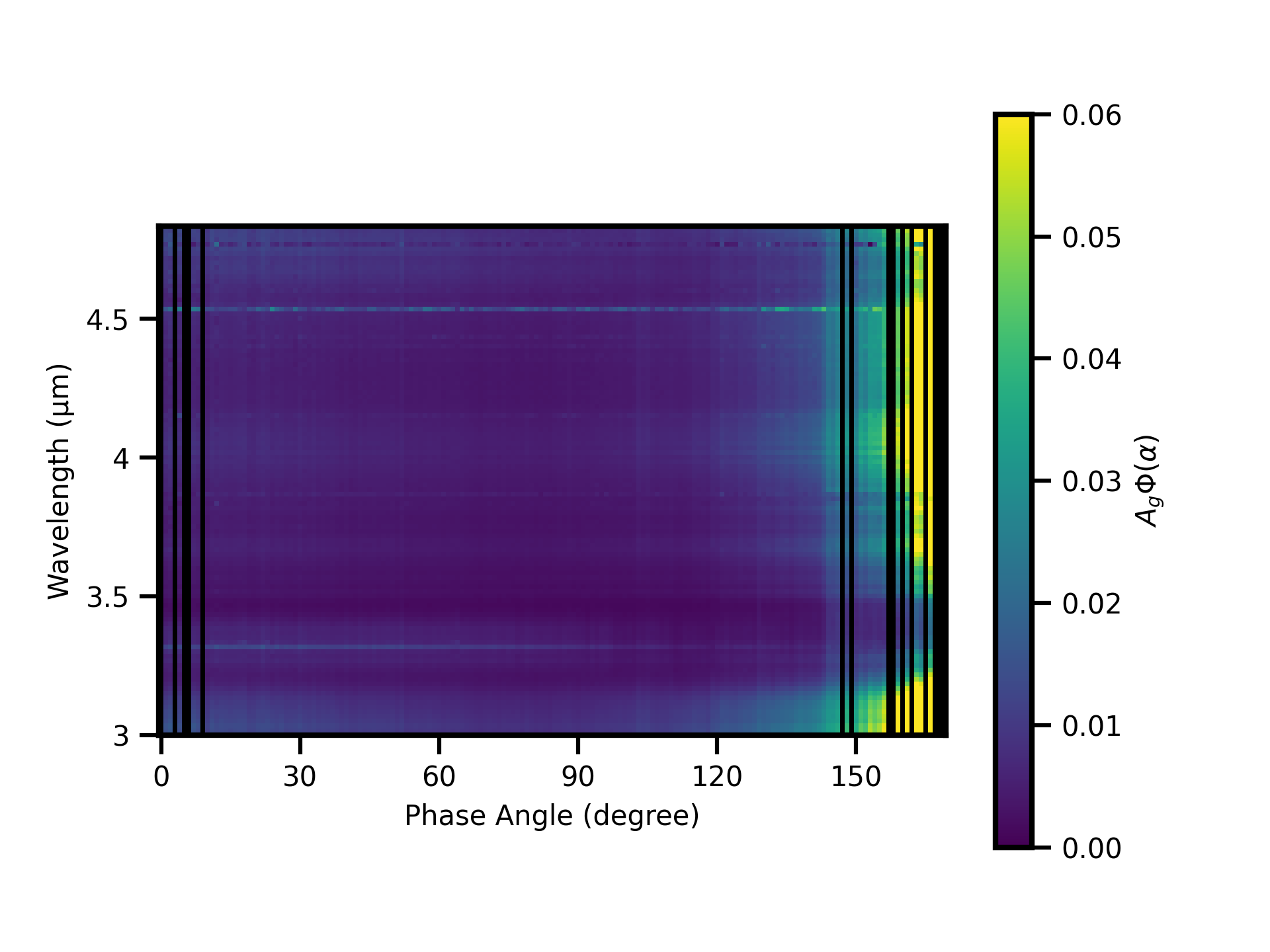}
    \caption{Non-local thermodynamic equilibrium emission signatures in the spectrally-resolved phase curves of Titan.  As explained in the text, the feature at 4.54\,$\upmu$m is likely affected by issues related to the instrument or cube calibration.}
    \label{fig:ch3d}
\end{figure}

Notable features in apparent emission are seen in Figure~\ref{fig:labelled} at 3.31\,$\upmu$m and 4.54\,$\upmu$m. For context, Figure~\ref{fig:emission_img} shows two high-quality examples of whole disk images of Titan at the wavelengths with these notable emission features. The 3.31\,$\upmu$m emission primarily comes from the dayside limb while the 4.54\,$\upmu$m emission is structured and concentrated near the southern pole. Figure~\ref{fig:ch3d} highlights the spectral region encompassing these features from the color-contour diagram in Figure~\ref{fig:heatmap}, demonstrating how both features maintain a near-constant $A_{\rm g}\Phi(\alpha)$ with increasing phase angle, which further suggests this feature is due to emission rather than phase-dependent forward scattering. \citet{2011Icar..214..571G} explain that the 3.31\,$\upmu$m feature is non-local thermodynamic equilibrium (non-LTE) emission from upper-atmospheric methane driven by absorption of solar radiation. The isotropic emission from this non-LTE source results in a phase curve shape that is distinctly less-structured than at wavelengths dominated by absorption or scattering processes, as seen in Figure~\ref{fig:heatmap}. This emission is prevalent enough to produce a spectral feature even when averaged over the disk of Titan. While both CO and CH$_3$D contribute to non-LTE emission near 4.54\,$\upmu$m, where the $\nu_2$ fundamental of CH$_3$D gives rise to a sharp feature at 4.54\,$\upmu$m \citep{2006P&SS...54.1552B}, emission from these species is not strong enough to explain a sharp increase in $A_{\rm g}\Phi(\alpha)$ seen in some cubes at this wavelength. Thus we suspect that a subset of cubes are affected by residual instrument or calibration issues at this wavelength, such as an order-sorting filter change, hot pixel, or other impediment in channel 317. 

\begin{figure}[b]
    \centering
    \includegraphics[width=\linewidth]{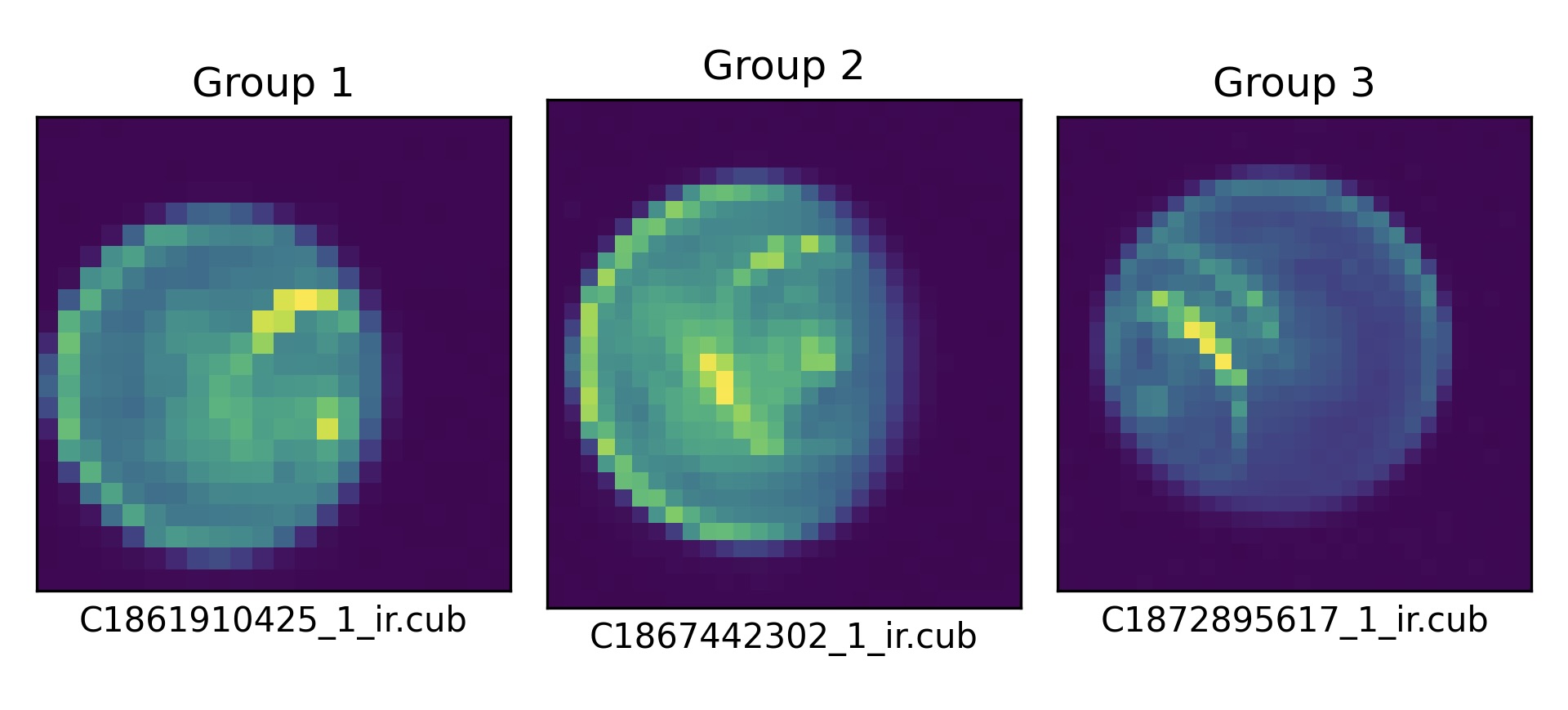}
    \caption{Images from identified ``bright'' flyby groups displaying likely cloud structures. Images are taken at 2.8\,$\upmu$m.}
    \label{fig:clouds}
\end{figure}

\subsection{Phase Curve Variability} \label{subsec:disc_split}

\begin{figure}[t]
    \centering
    \includegraphics[width=\linewidth]{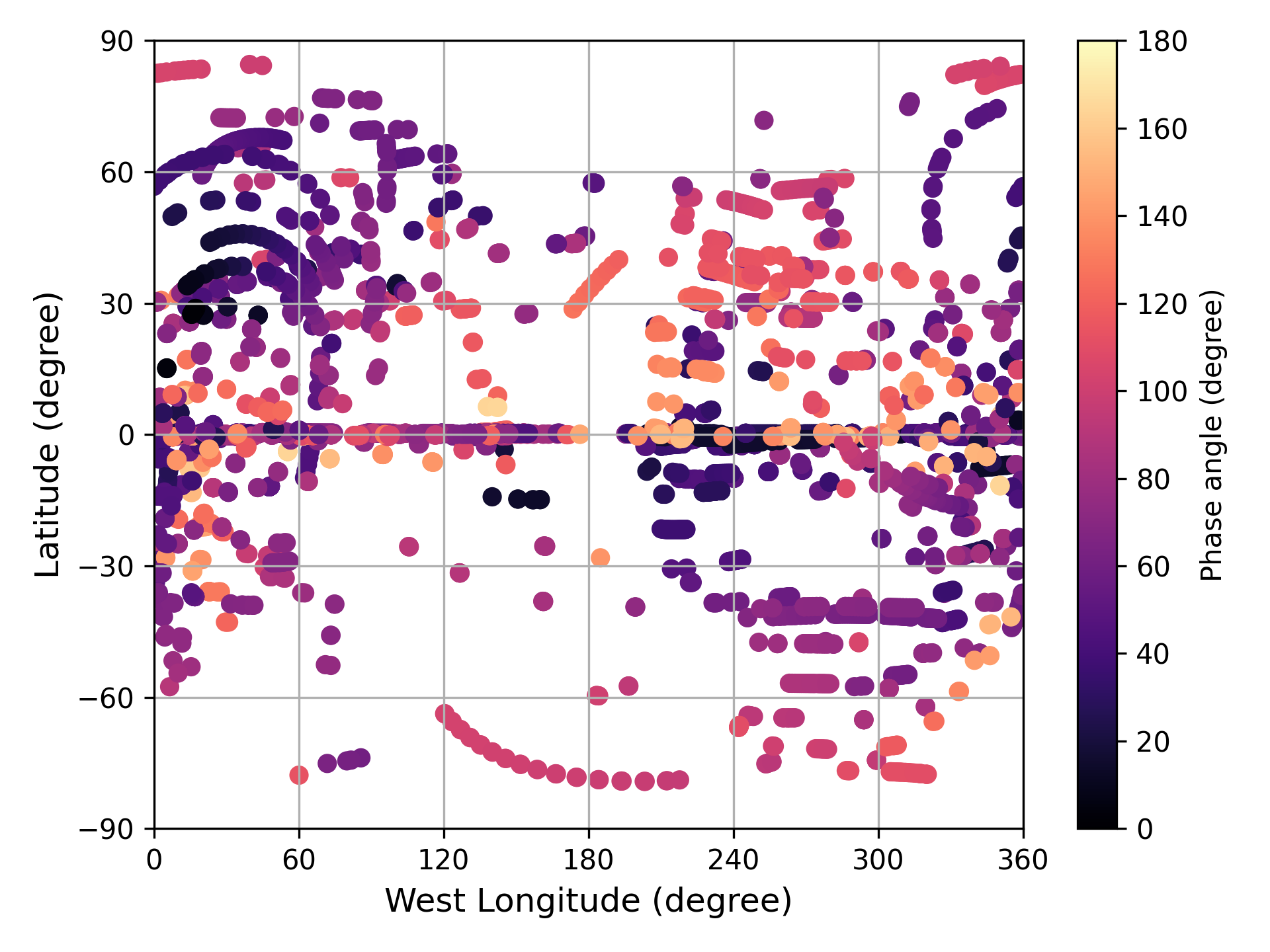}
    \caption{The spatial distribution of cubes by their sub-spacecraft coordinates. Most cubes used in the present study were acquired in quick succession by \textit{Cassini} during Titan passes, hence the clustering.}
    \label{fig:cubedistr}
\end{figure}

The 2\,$\upmu$m phase curve in Figure~\ref{fig:quadphasecurves} reveals a spread whose strength increases with decreasing phase angle. The spread far exceeds the anticipated error in disk-averaged values, especially at these longer wavelengths that do not suffer from saturation issues. This behavior is apparent in phase curves from all wavelengths with deep atmosphere and/or surface sensitivity, thus indicating that processes in the deep atmosphere or on the surface are driving the phase curve variability. For example, the visibility of Xanadu\,---\,a large, spectrally distinct region on the surface of Titan centered at about 100$\degree$S longitude and 10$\degree$S latitude \citep{Coustenis_1995,Lellouch_2004,Negrao_2006}\,---\,should introduce variability at surface-sensitive wavelengths. However, we found no substantive correlation between the sub-spacecraft coordinates of \textit{Cassini} at the time of cube acquisitions and the relative $A_{\rm g}\Phi(\alpha)$ of the cubes (Figure~\ref{fig:cubedistr}).

An indication that weather plays some role in causing variability in phase curves with deep atmospheric sensitivity is highlighted by flybys 253TI-255TI, flyby 264TI, and flyby 273TI (hereafter referred to as groups 1, 2, and 3, respectively), which dominate the brightest, low-phase observations in the 2\,$\upmu$m curve in Figure~\ref{fig:quadphasecurves}. Looking at visualizations of cubes from each group reveals a set of clouds above Titan’s north pole, surrounded by a ring of circumpolar clouds at about 50$\degree$N (Figure~\ref{fig:clouds}). All three groups took place during early- to mid-spring 2017. The cloud ring is most easily viewed in early group 3 cubes, while the polar cloud can easily be seen in cubes from groups 1 and 2. The location and shapes of cloud structures are in agreement with the findings of \citet{Yahn_2025}, who discovered that, from September 2016 to September 2017, clouds on Titan were concentrated around 0-120$\degree$W longitude, 50$\degree$N latitude, and had large aspect ratios indicating long, thin structures.

% \subsection{Application to Exoplanetary Atmospheres} \label{subsec:disc_exo}
\subsection{Applications and Future Work} \label{subsec:disc_exo}

The shape of Titan's phase curves are directly related to the physical properties of its hazes, such as their size, opacities, and single scattering albedo \citep{Tomasko_2008,Doose_2016,GarciaMunoz_2017}. Their wavelength-properties have been deduced previously from similar measurements at visible and some near-infrared wavelengths. The findings of this work could be used to extend the results of past studies by further constraining haze parameters at near-infrared wavelengths. This, in turn, would find applications in determining Titan's energy budget, modeling its atmospheric thermodynamics, or validating models thereof \citep{GarciaMunoz_2017}.

The presence of aerosols in Titan's atmosphere evidently dominates near-infrared reflected-light phase curves. It is reasonable, then, to expect similar mechanisms to occur in the atmospheres of hazy exoplanets. Models \citep{Hu_2013,Adams_2019,Gao_2020} and laboratory results \citep{He_2018,Horst_2018} have shown that exoplanets with a wide range of atmospheric conditions are capable of hosting aerosols, and many strong detections of aerosols come from warm/hot giant planets \citep[e.g.][]{Estrela_2021,Malsky_2025}. Observations of Titan can serve as a rare opportunity to study the phase curves of cool, terrestrial exoplanets with thick atmospheres, and especially the impact of atmospheric aerosols thereon.

The identification and study of aerosol forward-scattering can also avoid false positive detections of glint, whose strong specular scattering in the forward direction has been proposed as an avenue towards surface ocean detection \citep{Williams_2008,Robinson_2010,Vaughn_2023}. Misidentification of atmospheric forward scattering as glint could lead to an ocean false positive detection as both phenomena contribute at larger phase angles. Fortunately, Figure~\ref{fig:earthcomp} shows that, at least for Titan-like hazes, such false positive scenarios are unlikely. Importantly, while ocean glint increases Earth's crescent phase brightness by as much as 50\% at these wavelengths \citep{Robinson_2010}, this glint enhancement is markedly smaller than haze forward scattering effects in Titan's phase curve. Distinguishing Titan haze-like effects for an analogous exoplanet from glint requires access to only modest crescent phases (110--120$\degree$), where the Titan phase curve has a characteristic inflection point not seen in the Earth data. Phase-dependent color-color comparisons of Earth and Titan at red/NIR wavelengths, as shown in Figure~\ref{fig:colcol}, also demonstrate that, except at extreme crescent phases, brightness measurements at a few continuum wavelengths well-separates Titan-like worlds from potential exo-Earths.

\section{Conclusions} \label{sec:conclusion}

In this work, we used disk-averaged measurements of Titan's $A_{\rm g}\Phi(\alpha)$ to produce the first reflected-light phase curves of the moon that span the full near-infrared wavelength range. Key features of these phase curves are tied to atmospheric properties of Titan, and an understanding of these relations will be applicable to future efforts to directly image rocky exoplanets. Our key findings are as follows:

\begin{itemize}

    \item We developed a pipeline that takes \textit{Cassini} images of Titan and automatically determines the disk-averaged $A_{\rm g}\Phi(\alpha)$ of Titan at all VIMS-IR bands. Our dataset of 4,492 images spanned a phase angle range of 2$\degree$ to 165$\degree$. We produced phase curves of Titan in the near-infrared regime (0.88--5.12\,$\upmu$m) that show strong non-Lambertian effects.
    
    \item Our near-infrared reflectance spectrum of Titan at 2$\degree$ is a useful approximation to Titan's geometric albedo. This observation extends important previous results at primarily optical wavelengths \citep{Karkoschka_1998} and also agrees with very recent, high-quality near-full phase $A_{\rm g}\Phi(\alpha)$ spectral observations at near-infrared and infrared wavelengths from \textit{James Webb Space Telescope} \citep[][their Figure 4b]{Nixon_2025}.

    \item Phase curves of Titan at wavelengths with surface sensitivity show enhanced backscattering at low phase angles as well as increased variability at these phase angles due to clouds in the lower/deep atmosphere. Additionally, spectra show evidence of a feature at 3.31\,$\upmu$m attributable to CH$_4$ non-LTE emission. While this feature has been observed and studied before in spatially-resolved observations \citep{Baines_2005,2011Icar..214..571G}, we found that this feature is observable even when averaged over the entire disk.
    
    \item Phase curves of Titan at most wavelengths are dominated by forward scattering due to atmospheric aerosols. At all investigated wavelengths above about 1.3\,$\upmu$m, forward scattering by hazes causes Titan's disk-averaged $A_{\rm g}\Phi(\alpha)$ to be greater at crescent phase than at near-full phase. At certain wavelengths with significant atmospheric absorption, crescent phase $A_{\rm g}\Phi(\alpha)$ can be an order of magnitude greater than at near-full phase (Figure \ref{fig:curves_peaks}).
    
    \item  Future reflected light direct imaging of rocky exoplanets would yield phase curves for these distant worlds, and a detailed understanding of forward scattering effects on phase curves can reduce the chances of false positive detections of other phenomena (e.g., ocean glint) that occur at similar phase angles.
    
\end{itemize}

\section{Acknowledgments}
The present work was initially supported by a seed grant from the Arizona Astrobiology Center, led by CC. All authors acknowledge support through an award from NASA's Exoplanets Research Program (No.\,80NSSC25K7149). CC and TR conceived of this study, which was further developed by CC, TR, JB, and LM. All pipeline materials, results, and figures were created by CC. CC wrote all early drafts of this manuscript, with later editing support from TR, JB, and LM. We thank A.\,Garc\'ia Mu\~noz for sharing optical phase curve data for Titan. We also thank the two anonymous reviewers for their insightful comments and suggestions for improvement.

\begin{software}
    \newline astropy \citep{Astropy1,Astropy2,Astropy3}, ISIS3 \citep{Rodriguez2024}, matplotlib \citep{Hunter_2007}, numpy \citep{2020NumPy-Array}, pyvims \citep{LeMoulic_2019}, scipy \citep{2020SciPy-NMeth}
\end{software}

\section{Appendix - Data Acquisition and Calibration}
Data used in this work were acquired using the \href{https://pds-imaging.jpl.nasa.gov/search/}{PDS Image Atlas}. The data were part of the "VIMS
Observations from the Cassini Tour of the Saturn System" dataset \citep{vims2020}. Data were selected from the dataset by requiring that they target Titan. The ISIS3 package is developed by the United States Geological Survey and can be accessed from their \href{https://github.com/DOI-USGS/ISIS3?tab=readme-ov-file}{GitHub page}.

\newpage

\bibliography{biblist}{}

\begin{thebibliography}{}
\expandafter\ifx\csname natexlab\endcsname\relax\def\natexlab#1{#1}\fi
\providecommand{\url}[1]{\href{#1}{#1}}
\providecommand{\dodoi}[1]{doi:~\href{http://doi.org/#1}{\nolinkurl{#1}}}
\providecommand{\doeprint}[1]{\href{http://ascl.net/#1}{\nolinkurl{http://ascl.net/#1}}}
\providecommand{\doarXiv}[1]{\href{https://arxiv.org/abs/#1}{\nolinkurl{https://arxiv.org/abs/#1}}}

\bibitem[{Acton {et~al.}(2018)Acton, Bachman, Semenov, \& Wright}]{ACTON20189}
Acton, C., Bachman, N., Semenov, B., \& Wright, E. 2018, Planetary and Space Science, 150, 9, \dodoi{https://doi.org/10.1016/j.pss.2017.02.013}

\bibitem[{Acton(1996)}]{ACTON199665}
Acton, C.~H. 1996, Planetary and Space Science, 44, 65, \dodoi{https://doi.org/10.1016/0032-0633(95)00107-7}

\bibitem[{{Adams} {et~al.}(2019){Adams}, {Gao}, {de Pater}, \& {Morley}}]{Adams_2019}
{Adams}, D., {Gao}, P., {de Pater}, I., \& {Morley}, C.~V. 2019, \apj, 874, 61, \dodoi{10.3847/1538-4357/ab074c}

\bibitem[{{Arking} \& {Potter}(1968)}]{Arking_1968}
{Arking}, A., \& {Potter}, J. 1968, Journal of Atmospheric Sciences, 25, 617 , \dodoi{10.1175/1520-0469(1968)025<0617:TPCOVA>2.0.CO;2}

\bibitem[{{Astropy Collaboration} {et~al.}(2013){Astropy Collaboration}, {Robitaille}, {Tollerud}, {Greenfield}, {Droettboom}, {Bray}, {Aldcroft}, {Davis}, {Ginsburg}, {Price-Whelan}, {Kerzendorf}, {Conley}, {Crighton}, {Barbary}, {Muna}, {Ferguson}, {Grollier}, {Parikh}, {Nair}, {Unther}, {Deil}, {Woillez}, {Conseil}, {Kramer}, {Turner}, {Singer}, {Fox}, {Weaver}, {Zabalza}, {Edwards}, {Azalee Bostroem}, {Burke}, {Casey}, {Crawford}, {Dencheva}, {Ely}, {Jenness}, {Labrie}, {Lim}, {Pierfederici}, {Pontzen}, {Ptak}, {Refsdal}, {Servillat}, \& {Streicher}}]{Astropy1}
{Astropy Collaboration}, {Robitaille}, T.~P., {Tollerud}, E.~J., {et~al.} 2013, \aap, 558, A33, \dodoi{10.1051/0004-6361/201322068}

\bibitem[{{Astropy Collaboration} {et~al.}(2018){Astropy Collaboration}, {Price-Whelan}, {Sip{\H{o}}cz}, {G{\"u}nther}, {Lim}, {Crawford}, {Conseil}, {Shupe}, {Craig}, {Dencheva}, {Ginsburg}, {VanderPlas}, {Bradley}, {P{\'e}rez-Su{\'a}rez}, {de Val-Borro}, {Aldcroft}, {Cruz}, {Robitaille}, {Tollerud}, {Ardelean}, {Babej}, {Bach}, {Bachetti}, {Bakanov}, {Bamford}, {Barentsen}, {Barmby}, {Baumbach}, {Berry}, {Biscani}, {Boquien}, {Bostroem}, {Bouma}, {Brammer}, {Bray}, {Breytenbach}, {Buddelmeijer}, {Burke}, {Calderone}, {Cano Rodr{\'\i}guez}, {Cara}, {Cardoso}, {Cheedella}, {Copin}, {Corrales}, {Crichton}, {D'Avella}, {Deil}, {Depagne}, {Dietrich}, {Donath}, {Droettboom}, {Earl}, {Erben}, {Fabbro}, {Ferreira}, {Finethy}, {Fox}, {Garrison}, {Gibbons}, {Goldstein}, {Gommers}, {Greco}, {Greenfield}, {Groener}, {Grollier}, {Hagen}, {Hirst}, {Homeier}, {Horton}, {Hosseinzadeh}, {Hu}, {Hunkeler}, {Ivezi{\'c}}, {Jain}, {Jenness}, {Kanarek}, {Kendrew}, {Kern}, {Kerzendorf}, {Khvalko}, {King}, {Kirkby}, {Kulkarni},
  {Kumar}, {Lee}, {Lenz}, {Littlefair}, {Ma}, {Macleod}, {Mastropietro}, {McCully}, {Montagnac}, {Morris}, {Mueller}, {Mumford}, {Muna}, {Murphy}, {Nelson}, {Nguyen}, {Ninan}, {N{\"o}the}, {Ogaz}, {Oh}, {Parejko}, {Parley}, {Pascual}, {Patil}, {Patil}, {Plunkett}, {Prochaska}, {Rastogi}, {Reddy Janga}, {Sabater}, {Sakurikar}, {Seifert}, {Sherbert}, {Sherwood-Taylor}, {Shih}, {Sick}, {Silbiger}, {Singanamalla}, {Singer}, {Sladen}, {Sooley}, {Sornarajah}, {Streicher}, {Teuben}, {Thomas}, {Tremblay}, {Turner}, {Terr{\'o}n}, {van Kerkwijk}, {de la Vega}, {Watkins}, {Weaver}, {Whitmore}, {Woillez}, {Zabalza}, \& {Astropy Contributors}}]{Astropy2}
{Astropy Collaboration}, {Price-Whelan}, A.~M., {Sip{\H{o}}cz}, B.~M., {et~al.} 2018, \aj, 156, 123, \dodoi{10.3847/1538-3881/aabc4f}

\bibitem[{{Astropy Collaboration} {et~al.}(2022){Astropy Collaboration}, {Price-Whelan}, {Lim}, {Earl}, {Starkman}, {Bradley}, {Shupe}, {Patil}, {Corrales}, {Brasseur}, {N{\"o}the}, {Donath}, {Tollerud}, {Morris}, {Ginsburg}, {Vaher}, {Weaver}, {Tocknell}, {Jamieson}, {van Kerkwijk}, {Robitaille}, {Merry}, {Bachetti}, {G{\"u}nther}, {Aldcroft}, {Alvarado-Montes}, {Archibald}, {B{\'o}di}, {Bapat}, {Barentsen}, {Baz{\'a}n}, {Biswas}, {Boquien}, {Burke}, {Cara}, {Cara}, {Conroy}, {Conseil}, {Craig}, {Cross}, {Cruz}, {D'Eugenio}, {Dencheva}, {Devillepoix}, {Dietrich}, {Eigenbrot}, {Erben}, {Ferreira}, {Foreman-Mackey}, {Fox}, {Freij}, {Garg}, {Geda}, {Glattly}, {Gondhalekar}, {Gordon}, {Grant}, {Greenfield}, {Groener}, {Guest}, {Gurovich}, {Handberg}, {Hart}, {Hatfield-Dodds}, {Homeier}, {Hosseinzadeh}, {Jenness}, {Jones}, {Joseph}, {Kalmbach}, {Karamehmetoglu}, {Ka{\l}uszy{\'n}ski}, {Kelley}, {Kern}, {Kerzendorf}, {Koch}, {Kulumani}, {Lee}, {Ly}, {Ma}, {MacBride}, {Maljaars}, {Muna}, {Murphy}, {Norman},
  {O'Steen}, {Oman}, {Pacifici}, {Pascual}, {Pascual-Granado}, {Patil}, {Perren}, {Pickering}, {Rastogi}, {Roulston}, {Ryan}, {Rykoff}, {Sabater}, {Sakurikar}, {Salgado}, {Sanghi}, {Saunders}, {Savchenko}, {Schwardt}, {Seifert-Eckert}, {Shih}, {Jain}, {Shukla}, {Sick}, {Simpson}, {Singanamalla}, {Singer}, {Singhal}, {Sinha}, {Sip{\H{o}}cz}, {Spitler}, {Stansby}, {Streicher}, {{\v{S}}umak}, {Swinbank}, {Taranu}, {Tewary}, {Tremblay}, {de Val-Borro}, {Van Kooten}, {Vasovi{\'c}}, {Verma}, {de Miranda Cardoso}, {Williams}, {Wilson}, {Winkel}, {Wood-Vasey}, {Xue}, {Yoachim}, {Zhang}, {Zonca}, \& {Astropy Project Contributors}}]{Astropy3}
{Astropy Collaboration}, {Price-Whelan}, A.~M., {Lim}, P.~L., {et~al.} 2022, \apj, 935, 167, \dodoi{10.3847/1538-4357/ac7c74}

\bibitem[{Baines {et~al.}(2005)Baines, Drossart, Momary, Formisano, Griffith, Bellucci, Bibring, Brown, Buratti, Capaccioni, {et~al.}}]{Baines_2005}
Baines, K.~H., Drossart, P., Momary, T.~W., {et~al.} 2005, Earth, Moon, and Planets, 96, 119, \dodoi{10.1007/s11038-005-9058-2}

\bibitem[{{Baines} {et~al.}(2006){Baines}, {Drossart}, {Lopez-Valverde}, {Atreya}, {Sotin}, {Momary}, {Brown}, {Buratti}, {Clark}, \& {Nicholson}}]{2006P&SS...54.1552B}
{Baines}, K.~H., {Drossart}, P., {Lopez-Valverde}, M.~A., {et~al.} 2006, \planss, 54, 1552, \dodoi{10.1016/j.pss.2006.06.020}

\bibitem[{Barnes {et~al.}(2014)Barnes, Sotin, Soderblom, Brown, Hayes, Donelan, Rodriguez, Le~Mou{\'e}lic, Baines, \& Mccord}]{barnes_2014}
Barnes, J., Sotin, C., Soderblom, J., {et~al.} 2014, {Planetary Science}, 3, 3, \dodoi{10.1186/s13535-014-0003-4}

\bibitem[{Barnes {et~al.}(2011)Barnes, Soderblom, Brown, Soderblom, Stephan, Jaumann, Mouélic, Rodriguez, Sotin, Buratti, Baines, Clark, \& Nicholson}]{Barnes_2011}
Barnes, J.~W., Soderblom, J.~M., Brown, R.~H., {et~al.} 2011, Icarus, 211, 722, \dodoi{https://doi.org/10.1016/j.icarus.2010.09.022}

\bibitem[{Barnes {et~al.}(2013)Barnes, Clark, Sotin, Ádámkovics, Appéré, Rodriguez, Soderblom, Brown, Buratti, Baines, Le~Mouélic, \& Nicholson}]{Barnes_2013}
Barnes, J.~W., Clark, R.~N., Sotin, C., {et~al.} 2013, The Astrophysical Journal, 777, 161, \dodoi{10.1088/0004-637X/777/2/161}

\bibitem[{{Brandeker} {et~al.}(2022){Brandeker}, {Heng}, {Lendl}, {Patel}, {Morris}, {Broeg}, {Guterman}, {Beck}, {Maxted}, {Demangeon}, {Delrez}, {Demory}, {Kitzmann}, {Santos}, {Singh}, {Alibert}, {Alonso}, {Anglada}, {B{\'a}rczy}, {Barrado y Navascues}, {Barros}, {Baumjohann}, {Beck}, {Benz}, {Billot}, {Bonfils}, {Bruno}, {Cabrera}, {Charnoz}, {Collier Cameron}, {Corral van Damme}, {Csizmadia}, {Davies}, {Deleuil}, {Deline}, {Ehrenreich}, {Erikson}, {Farinato}, {Fortier}, {Fossati}, {Fridlund}, {Gandolfi}, {Gillon}, {G{\"u}del}, {Hoyer}, {Isaak}, {Kiss}, {Laskar}, {Lecavelier des Etangs}, {Lovis}, {Luntzer}, {Magrin}, {Nascimbeni}, {Olofsson}, {Ottensamer}, {Pagano}, {Pall{\'e}}, {Peter}, {Piotto}, {Pollacco}, {Queloz}, {Ragazzoni}, {Rando}, {Rauer}, {Ribas}, {Scandariato}, {S{\'e}gransan}, {Simon}, {Smith}, {Sousa}, {Steller}, {Szab{\'o}}, {Thomas}, {Udry}, {Van Grootel}, {Walton}, \& {Wolter}}]{Brandeker_2022}
{Brandeker}, A., {Heng}, K., {Lendl}, M., {et~al.} 2022, \aap, 659, L4, \dodoi{10.1051/0004-6361/202243082}

\bibitem[{{Brown} \& {VIMS Science Team}(2020)}]{vims2020}
{Brown}, R.~H., \& {VIMS Science Team}. 2020, VIMS Observations from the Cassini Tour of the Saturn System, \dodoi{10.17189/1504134}

\bibitem[{Brown {et~al.}(2004)Brown, Baines, Bellucci, Bibring, Buratti, Capaccioni, Cerroni, Clark, Coradini, Cruikshank, \& et~al.}]{Brown_2004}
Brown, R.~H., Baines, K.~H., Bellucci, G., {et~al.} 2004, Space Science Reviews, 115, 111–168, \dodoi{10.1007/s11214-004-1453-x}

\bibitem[{{Buratti} \& {Veverka}(1984)}]{Buratti_1984}
{Buratti}, B., \& {Veverka}, J. 1984, \icarus, 58, 254, \dodoi{10.1016/0019-1035(84)90042-3}

\bibitem[{Canny(1986)}]{Canny_1986}
Canny, J. 1986, Pattern Analysis and Machine Intelligence, IEEE Transactions on, PAMI-8, 679 , \dodoi{10.1109/TPAMI.1986.4767851}

\bibitem[{Charbonneau {et~al.}(2005)Charbonneau, Allen, Megeath, Torres, Alonso, Brown, Gilliland, Latham, Mandushev, O’Donovan, \& Sozzetti}]{Charbonneau_2005}
Charbonneau, D., Allen, L.~E., Megeath, S.~T., {et~al.} 2005, The Astrophysical Journal, 626, 523, \dodoi{10.1086/429991}

\bibitem[{Coustenis {et~al.}(1995)Coustenis, Lellouch, Maillard, \& McKay}]{Coustenis_1995}
Coustenis, A., Lellouch, E., Maillard, J., \& McKay, C. 1995, Icarus, 118, 87, \dodoi{https://doi.org/10.1006/icar.1995.1179}

\bibitem[{{Doose} {et~al.}(2016){Doose}, {Karkoschka}, {Tomasko}, \& {Anderson}}]{Doose_2016}
{Doose}, L.~R., {Karkoschka}, E., {Tomasko}, M.~G., \& {Anderson}, C.~M. 2016, \icarus, 270, 355, \dodoi{10.1016/j.icarus.2015.09.039}

\bibitem[{Dyudina {et~al.}(2016)Dyudina, Zhang, Li, Kopparla, Ingersoll, Dones, Verbiscer, \& Yung}]{Dyudina_2016}
Dyudina, U., Zhang, X., Li, L., {et~al.} 2016, The Astrophysical Journal, 822, 76, \dodoi{10.3847/0004-637X/822/2/76}

\bibitem[{Es-sayeh {et~al.}(2023)Es-sayeh, Rodriguez, Coutelier, Rannou, Bézard, Maltagliati, Cornet, Grieger, Karkoschka, Le~Mouélic, Le~Gall, Neish, MacKenzie, Solomonidou, Sotin, \& Coustenis}]{Es-sayeh_2023}
Es-sayeh, M., Rodriguez, S., Coutelier, M., {et~al.} 2023, The Planetary Science Journal, 4, 44, \dodoi{10.3847/PSJ/acbd37}

\bibitem[{{Esteves} {et~al.}(2015){Esteves}, {De Mooij}, \& {Jayawardhana}}]{Esteves_2015}
{Esteves}, L.~J., {De Mooij}, E. J.~W., \& {Jayawardhana}, R. 2015, \apj, 804, 150, \dodoi{10.1088/0004-637X/804/2/150}

\bibitem[{{Estrela} {et~al.}(2021){Estrela}, {Swain}, {Roudier}, {West}, {Sedaghati}, \& {Valio}}]{Estrela_2021}
{Estrela}, R., {Swain}, M.~R., {Roudier}, G.~M., {et~al.} 2021, \aj, 162, 91, \dodoi{10.3847/1538-3881/ac0c7c}

\bibitem[{Feinberg {et~al.}(2024)Feinberg, Ziemer, Ansdell, Crooke, Dressing, Mennesson, O'Meara, Pepper, \& Roberge}]{feinbergetal2024}
Feinberg, L., Ziemer, J., Ansdell, M., {et~al.} 2024, in Space Telescopes and Instrumentation 2024: Optical, Infrared, and Millimeter Wave, ed. L.~E. Coyle, S.~Matsuura, \& M.~D. Perrin, Vol. 13092, International Society for Optics and Photonics (SPIE), 130921N, \dodoi{10.1117/12.3018328}

\bibitem[{{Gao} {et~al.}(2020){Gao}, {Thorngren}, {Lee}, {Fortney}, {Morley}, {Wakeford}, {Powell}, {Stevenson}, \& {Zhang}}]{Gao_2020}
{Gao}, P., {Thorngren}, D.~P., {Lee}, E. K.~H., {et~al.} 2020, Nature Astronomy, 4, 951, \dodoi{10.1038/s41550-020-1114-3}

\bibitem[{{Garc{\'\i}a-Comas} {et~al.}(2011){Garc{\'\i}a-Comas}, {L{\'o}pez-Puertas}, {Funke}, {Dinelli}, {Luisa Moriconi}, {Adriani}, {Molina}, \& {Coradini}}]{2011Icar..214..571G}
{Garc{\'\i}a-Comas}, M., {L{\'o}pez-Puertas}, M., {Funke}, B., {et~al.} 2011, \icarus, 214, 571, \dodoi{10.1016/j.icarus.2011.03.020}

\bibitem[{{Garc{\'\i}a Mu{\~n}oz} {et~al.}(2017){Garc{\'\i}a Mu{\~n}oz}, {Lavvas}, \& {West}}]{GarciaMunoz_2017}
{Garc{\'\i}a Mu{\~n}oz}, A., {Lavvas}, P., \& {West}, R.~A. 2017, Nature Astronomy, 1, 0114, \dodoi{10.1038/s41550-017-0114}

\bibitem[{Harris {et~al.}(2020)Harris, Millman, van~der Walt, Gommers, Virtanen, Cournapeau, Wieser, Taylor, Berg, Smith, Kern, Picus, Hoyer, van Kerkwijk, Brett, Haldane, Fernández~del Río, Wiebe, Peterson, Gérard-Marchant, Sheppard, Reddy, Weckesser, Abbasi, Gohlke, \& Oliphant}]{2020NumPy-Array}
Harris, C.~R., Millman, K.~J., van~der Walt, S.~J., {et~al.} 2020, Nature, 585, 357–362, \dodoi{10.1038/s41586-020-2649-2}

\bibitem[{{He} {et~al.}(2018){He}, {H{\"o}rst}, {Lewis}, {Yu}, {Moses}, {Kempton}, {Marley}, {McGuiggan}, {Morley}, {Valenti}, \& {Vuitton}}]{He_2018}
{He}, C., {H{\"o}rst}, S.~M., {Lewis}, N.~K., {et~al.} 2018, \aj, 156, 38, \dodoi{10.3847/1538-3881/aac883}

\bibitem[{Hedman \& Stark(2015)}]{Hedman_2015}
Hedman, M.~M., \& Stark, C.~C. 2015, The Astrophysical Journal, 811, 67, \dodoi{10.1088/0004-637X/811/1/67}

\bibitem[{{Heng} \& {Li}(2021)}]{Heng_2021}
{Heng}, K., \& {Li}, L. 2021, \apjl, 909, L20, \dodoi{10.3847/2041-8213/abe872}

\bibitem[{{Hoeijmakers} {et~al.}(2018){Hoeijmakers}, {Snellen}, \& {van Terwisga}}]{Hoeijmakers_2018}
{Hoeijmakers}, H.~J., {Snellen}, I.~A.~G., \& {van Terwisga}, S.~E. 2018, \aap, 610, A47, \dodoi{10.1051/0004-6361/201731192}

\bibitem[{{Hooton} {et~al.}(2022){Hooton}, {Hoyer}, {Kitzmann}, {Morris}, {Smith}, {Collier Cameron}, {Futyan}, {Maxted}, {Queloz}, {Demory}, {Heng}, {Lendl}, {Cabrera}, {Csizmadia}, {Deline}, {Parviainen}, {Salmon}, {Sulis}, {Wilson}, {Bonfanti}, {Brandeker}, {Demangeon}, {Oshagh}, {Persson}, {Scandariato}, {Alibert}, {Alonso}, {Anglada Escud{\'e}}, {B{\'a}rczy}, {Barrado}, {Barros}, {Baumjohann}, {Beck}, {Beck}, {Benz}, {Billot}, {Bonfils}, {Bourrier}, {Broeg}, {Busch}, {Charnoz}, {Davies}, {Deleuil}, {Delrez}, {Ehrenreich}, {Erikson}, {Farinato}, {Fortier}, {Fossati}, {Fridlund}, {Gandolfi}, {Gillon}, {G{\"u}del}, {Isaak}, {Jones}, {Kiss}, {Laskar}, {Lecavelier des Etangs}, {Lovis}, {Luntzer}, {Magrin}, {Nascimbeni}, {Olofsson}, {Ottensamer}, {Pagano}, {Pall{\'e}}, {Peter}, {Piotto}, {Pollacco}, {Ragazzoni}, {Rando}, {Ratti}, {Rauer}, {Ribas}, {Santos}, {S{\'e}gransan}, {Simon}, {Sousa}, {Steller}, {Szab{\'o}}, {Thomas}, {Udry}, {Ulmer}, {Van Grootel}, \& {Walton}}]{Hooton_2022}
{Hooton}, M.~J., {Hoyer}, S., {Kitzmann}, D., {et~al.} 2022, \aap, 658, A75, \dodoi{10.1051/0004-6361/202141645}

\bibitem[{{H{\"o}rst} {et~al.}(2018){H{\"o}rst}, {He}, {Lewis}, {Kempton}, {Marley}, {Morley}, {Moses}, {Valenti}, \& {Vuitton}}]{Horst_2018}
{H{\"o}rst}, S.~M., {He}, C., {Lewis}, N.~K., {et~al.} 2018, Nature Astronomy, 2, 303, \dodoi{10.1038/s41550-018-0397-0}

\bibitem[{Hu {et~al.}(2013)Hu, Seager, \& Bains}]{Hu_2013}
Hu, R., Seager, S., \& Bains, W. 2013, The Astrophysical Journal, 769, 6, \dodoi{10.1088/0004-637X/769/1/6}

\bibitem[{Hunter(2007)}]{Hunter_2007}
Hunter, J.~D. 2007, Computing in Science \& Engineering, 9, 90, \dodoi{10.1109/MCSE.2007.55}

\bibitem[{{Karkoschka}(1998)}]{Karkoschka_1998}
{Karkoschka}, E. 1998, \icarus, 133, 134, \dodoi{10.1006/icar.1998.5913}

\bibitem[{{Kelvin Rodriguez} \& {Astrogeology Science Center}(2024)}]{isis}
{Kelvin Rodriguez}, \& {Astrogeology Science Center}. 2024, Integrated Software for Imagers and Spectrometers (ISIS) 8.3.0,  U.S. Geological Survey, \dodoi{10.5066/P13TADS5}

\bibitem[{{Kesseli} {et~al.}(2022){Kesseli}, {Snellen}, {Casasayas-Barris}, {Molli{\`e}re}, \& {S{\'a}nchez-L{\'o}pez}}]{Kesseli_2022}
{Kesseli}, A.~Y., {Snellen}, I.~A.~G., {Casasayas-Barris}, N., {Molli{\`e}re}, P., \& {S{\'a}nchez-L{\'o}pez}, A. 2022, \aj, 163, 107, \dodoi{10.3847/1538-3881/ac4336}

\bibitem[{{Knutson} {et~al.}(2008){Knutson}, {Charbonneau}, {Allen}, {Burrows}, \& {Megeath}}]{Knutson_2006}
{Knutson}, H.~A., {Charbonneau}, D., {Allen}, L.~E., {Burrows}, A., \& {Megeath}, S.~T. 2008, \apj, 673, 526, \dodoi{10.1086/523894}

\bibitem[{{Kreidberg} {et~al.}(2014){Kreidberg}, {Bean}, {D{\'e}sert}, {Line}, {Fortney}, {Madhusudhan}, {Stevenson}, {Showman}, {Charbonneau}, {McCullough}, {Seager}, {Burrows}, {Henry}, {Williamson}, {Kataria}, \& {Homeier}}]{Kreidberg_2014}
{Kreidberg}, L., {Bean}, J.~L., {D{\'e}sert}, J.-M., {et~al.} 2014, \apjl, 793, L27, \dodoi{10.1088/2041-8205/793/2/L27}

\bibitem[{Lavvas {et~al.}(2010)Lavvas, Yelle, \& Griffith}]{LAVVAS2010832}
Lavvas, P., Yelle, R., \& Griffith, C. 2010, Icarus, 210, 832, \dodoi{https://doi.org/10.1016/j.icarus.2010.07.025}

\bibitem[{{Le Mouélic} {et~al.}(2019){Le Mouélic}, Cornet, Rodriguez, Sotin, Seignovert, Barnes, Brown, Baines, Buratti, Clark, Nicholson, Lasue, Pasek, \& Soderblom}]{LeMoulic_2019}
{Le Mouélic}, S., Cornet, T., Rodriguez, S., {et~al.} 2019, Icarus, 319, 121, \dodoi{https://doi.org/10.1016/j.icarus.2018.09.017}

\bibitem[{Lellouch {et~al.}(2004)Lellouch, Schmitt, Coustenis, \& Cuby}]{Lellouch_2004}
Lellouch, E., Schmitt, B., Coustenis, A., \& Cuby, J.-G. 2004, Icarus, 168, 209, \dodoi{https://doi.org/10.1016/j.icarus.2003.12.001}

\bibitem[{{Lendl} {et~al.}(2020){Lendl}, {Csizmadia}, {Deline}, {Fossati}, {Kitzmann}, {Heng}, {Hoyer}, {Salmon}, {Benz}, {Broeg}, {Ehrenreich}, {Fortier}, {Queloz}, {Bonfanti}, {Brandeker}, {Collier Cameron}, {Delrez}, {Garcia Mu{\~n}oz}, {Hooton}, {Maxted}, {Morris}, {Van Grootel}, {Wilson}, {Alibert}, {Alonso}, {Asquier}, {Bandy}, {B{\'a}rczy}, {Barrado}, {Barros}, {Baumjohann}, {Beck}, {Beck}, {Bekkelien}, {Bergomi}, {Billot}, {Biondi}, {Bonfils}, {Bourrier}, {Busch}, {Cabrera}, {Cessa}, {Charnoz}, {Chazelas}, {Corral Van Damme}, {Davies}, {Deleuil}, {Demangeon}, {Demory}, {Erikson}, {Farinato}, {Fridlund}, {Futyan}, {Gandolfi}, {Gillon}, {Guterman}, {Hasiba}, {Hernandez}, {Isaak}, {Kiss}, {Kuntzer}, {Lecavelier des Etangs}, {L{\"u}ftinger}, {Laskar}, {Lovis}, {Magrin}, {Malvasio}, {Marafatto}, {Michaelis}, {Munari}, {Nascimbeni}, {Olofsson}, {Ottacher}, {Ottensamer}, {Pagano}, {Pall{\'e}}, {Peter}, {Piazza}, {Piotto}, {Pollacco}, {Ratti}, {Rauer}, {Ragazzoni}, {Rando}, {Ribas}, {Rieder}, {Rohlfs},
  {Safa}, {Santos}, {Scandariato}, {S{\'e}gransan}, {Simon}, {Singh}, {Smith}, {Sordet}, {Sousa}, {Steller}, {Szab{\'o}}, {Thomas}, {Tschentscher}, {Udry}, {Viotto}, {Walter}, {Walton}, {Wildi}, \& {Wolter}}]{Lendl_2020}
{Lendl}, M., {Csizmadia}, S., {Deline}, A., {et~al.} 2020, \aap, 643, A94, \dodoi{10.1051/0004-6361/202038677}

\bibitem[{Madhusudhan {et~al.}(2020)Madhusudhan, Nixon, Welbanks, Piette, \& Booth}]{Madhusudhan_2020}
Madhusudhan, N., Nixon, M.~C., Welbanks, L., Piette, A. A.~A., \& Booth, R.~A. 2020, The Astrophysical Journal Letters, 891, L7, \dodoi{10.3847/2041-8213/ab7229}

\bibitem[{Madhusudhan {et~al.}(2023)Madhusudhan, Sarkar, Constantinou, Holmberg, Piette, \& Moses}]{Madhusudhan_2023}
Madhusudhan, N., Sarkar, S., Constantinou, S., {et~al.} 2023, The Astrophysical Journal Letters, 956, L13, \dodoi{10.3847/2041-8213/acf577}

\bibitem[{{Malsky} {et~al.}(2025){Malsky}, {Rauscher}, {Stevenson}, {Savel}, {Steinrueck}, {Gao}, {Kempton}, {Roman}, {Bean}, {Zhang}, {Parmentier}, {Piette}, \& {Kataria}}]{Malsky_2025}
{Malsky}, I., {Rauscher}, E., {Stevenson}, K., {et~al.} 2025, \aj, 169, 221, \dodoi{10.3847/1538-3881/adb7e8}

\bibitem[{Mayorga {et~al.}(2016)Mayorga, Jackiewicz, Rages, West, Knowles, Lewis, \& Marley}]{Mayorga_2016}
Mayorga, L.~C., Jackiewicz, J., Rages, K., {et~al.} 2016, The Astronomical Journal, 152, 209, \dodoi{10.3847/0004-6256/152/6/209}

\bibitem[{Negrão {et~al.}(2006)Negrão, Coustenis, Lellouch, Maillard, Rannou, Schmitt, McKay, \& Boudon}]{Negrao_2006}
Negrão, A., Coustenis, A., Lellouch, E., {et~al.} 2006, Planetary and Space Science, 54, 1225, \dodoi{https://doi.org/10.1016/j.pss.2006.05.031}

\bibitem[{Nixon {et~al.}(2025)Nixon, Bézard, Cornet, Coy, de~Pater, Es-Sayeh, Hammel, Lellouch, Lombardo, López-Puertas, Lora, Rannou, Rodriguez, Teanby, Turtle, Achterberg, Alvarez, Davies, de~Kleer, Doppmann, Fletcher, Hayes, Holler, Irwin, Jordan, King, Kutsop, Marlin, Melin, Milam, Molter, Moore, Nyffenegger-Péré, O’Donoghue, O’Meara, Rafkin, Roman, Rostopchina, Rowe-Gurney, Schmidt, Schmidt, Sotin, Stallard, Stansberry, \& West}]{Nixon_2025}
Nixon, C.~A., Bézard, B., Cornet, T., {et~al.} 2025, Nature Astronomy, \dodoi{10.1038/s41550-025-02537-3}

\bibitem[{{Pall{\'e}} {et~al.}(2003){Pall{\'e}}, {Goode}, {Yurchyshyn}, {Qiu}, {Hickey}, {Monta{\~n}{\'e}S Rodriguez}, {Chu}, {Kolbe}, {Brown}, \& {Koonin}}]{Palle_2003}
{Pall{\'e}}, E., {Goode}, P.~R., {Yurchyshyn}, V., {et~al.} 2003, Journal of Geophysical Research (Atmospheres), 108, 4710, \dodoi{10.1029/2003JD003611}

\bibitem[{Parmentier \& Crossfield(2018)}]{Parmentier_2018}
Parmentier, V., \& Crossfield, I. J.~M. 2018, Exoplanet Phase Curves: Observations and Theory (Springer International Publishing), 1419–1440, \dodoi{10.1007/978-3-319-55333-7_116}

\bibitem[{{Qiu} {et~al.}(2003){Qiu}, {Goode}, {Pall{\'e}}, {Yurchyshyn}, {Hickey}, {Monta{\~n}{\'e}S Rodriguez}, {Chu}, {Kolbe}, {Brown}, \& {Koonin}}]{Qiu_2003}
{Qiu}, J., {Goode}, P.~R., {Pall{\'e}}, E., {et~al.} 2003, Journal of Geophysical Research (Atmospheres), 108, 4709, \dodoi{10.1029/2003JD003610}

\bibitem[{{Rages} {et~al.}(1983){Rages}, {Pollack}, \& {Smith}}]{Rages_1983}
{Rages}, K., {Pollack}, J.~B., \& {Smith}, P.~H. 1983, \jgr, 88, 8721, \dodoi{10.1029/JA088iA11p08721}

\bibitem[{{Robinson} {et~al.}(2014){Robinson}, {Maltagliati}, {Marley}, \& {Fortney}}]{Robinson_2014}
{Robinson}, T.~D., {Maltagliati}, L., {Marley}, M.~S., \& {Fortney}, J.~J. 2014, Proceedings of the National Academy of Science, 111, 9042, \dodoi{10.1073/pnas.1403473111}

\bibitem[{{Robinson} {et~al.}(2010){Robinson}, {Meadows}, \& {Crisp}}]{Robinson_2010}
{Robinson}, T.~D., {Meadows}, V.~S., \& {Crisp}, D. 2010, \apjl, 721, L67, \dodoi{10.1088/2041-8205/721/1/L67}

\bibitem[{Rodriguez \& {Astrogeology Science Center}(2024)}]{Rodriguez2024}
Rodriguez, K., \& {Astrogeology Science Center}. 2024, Integrated software for imagers and spectrometers ({ISIS}) 8.3.0,  U.S. Geological Survey

\bibitem[{{Rustamkulov} {et~al.}(2023){Rustamkulov}, {Sing}, {Mukherjee}, {May}, {Kirk}, {Schlawin}, {Line}, {Piaulet}, {Carter}, {Batalha}, {Goyal}, {L{\'o}pez-Morales}, {Lothringer}, {MacDonald}, {Moran}, {Stevenson}, {Wakeford}, {Espinoza}, {Bean}, {Batalha}, {Benneke}, {Berta-Thompson}, {Crossfield}, {Gao}, {Kreidberg}, {Powell}, {Cubillos}, {Gibson}, {Leconte}, {Molaverdikhani}, {Nikolov}, {Parmentier}, {Roy}, {Taylor}, {Turner}, {Wheatley}, {Aggarwal}, {Ahrer}, {Alam}, {Alderson}, {Allen}, {Banerjee}, {Barat}, {Barrado}, {Barstow}, {Bell}, {Blecic}, {Brande}, {Casewell}, {Changeat}, {Chubb}, {Crouzet}, {Daylan}, {Decin}, {D{\'e}sert}, {Mikal-Evans}, {Feinstein}, {Flagg}, {Fortney}, {Harrington}, {Heng}, {Hong}, {Hu}, {Iro}, {Kataria}, {Kempton}, {Krick}, {Lendl}, {Lillo-Box}, {Louca}, {Lustig-Yaeger}, {Mancini}, {Mansfield}, {Mayne}, {Miguel}, {Morello}, {Ohno}, {Palle}, {Petit dit de la Roche}, {Rackham}, {Radica}, {Ramos-Rosado}, {Redfield}, {Rogers}, {Shkolnik}, {Southworth}, {Teske}, {Tremblin},
  {Tucker}, {Venot}, {Waalkes}, {Welbanks}, {Zhang}, \& {Zieba}}]{Rustamkulov_2023}
{Rustamkulov}, Z., {Sing}, D.~K., {Mukherjee}, S., {et~al.} 2023, \nat, 614, 659, \dodoi{10.1038/s41586-022-05677-y}

\bibitem[{Soderblom {et~al.}(2012)Soderblom, Barnes, Soderblom, Brown, Griffith, Nicholson, Stephan, Jaumann, Sotin, Baines, Buratti, \& Clark}]{Soderblom_2012}
Soderblom, J.~M., Barnes, J.~W., Soderblom, L.~A., {et~al.} 2012, Icarus, 220, 744, \dodoi{https://doi.org/10.1016/j.icarus.2012.05.030}

\bibitem[{{Stephan} {et~al.}(2010){Stephan}, {Jaumann}, {Brown}, {Soderblom}, {Soderblom}, {Barnes}, {Sotin}, {Griffith}, {Kirk}, {Baines}, {Buratti}, {Clark}, {Lytle}, {Nelson}, \& {Nicholson}}]{Stephan_2010}
{Stephan}, K., {Jaumann}, R., {Brown}, R.~H., {et~al.} 2010, \grl, 37, L07104, \dodoi{10.1029/2009GL042312}

\bibitem[{{Stevenson} {et~al.}(2014){Stevenson}, {D{\'e}sert}, {Line}, {Bean}, {Fortney}, {Showman}, {Kataria}, {Kreidberg}, {McCullough}, {Henry}, {Charbonneau}, {Burrows}, {Seager}, {Madhusudhan}, {Williamson}, \& {Homeier}}]{Stevenson_2014}
{Stevenson}, K.~B., {D{\'e}sert}, J.-M., {Line}, M.~R., {et~al.} 2014, Science, 346, 838, \dodoi{10.1126/science.1256758}

\bibitem[{Strauss {et~al.}(2024)Strauss, Robinson, Trilling, Cummings, \& Smith}]{Strauss_2024}
Strauss, R.~H., Robinson, T.~D., Trilling, D.~E., Cummings, R., \& Smith, C.~J. 2024, The Astronomical Journal, 167, 87, \dodoi{10.3847/1538-3881/ad1bd1}

\bibitem[{{Titan Discipline Working Group}(2019)}]{CassiniTitan_2019}
{Titan Discipline Working Group}. 2019, {Titan}, Tech. rep., NASA

\bibitem[{{Tomasko} {et~al.}(2008){Tomasko}, {Doose}, {Engel}, {Dafoe}, {West}, {Lemmon}, {Karkoschka}, \& {See}}]{Tomasko_2008}
{Tomasko}, M.~G., {Doose}, L., {Engel}, S., {et~al.} 2008, \planss, 56, 669, \dodoi{10.1016/j.pss.2007.11.019}

\bibitem[{Vaughan {et~al.}(2023)Vaughan, Gebhard, Bott, Casewell, Cowan, Doelman, Kenworthy, Mazoyer, Millar-Blanchaer, Trees, Stam, Absil, Altinier, Baudoz, Belikov, Bidot, Birkby, Bonse, Brandl, Carlotti, Choquet, van Dam, Desai, Fogarty, Fowler, van Gorkom, Gutierrez, Guyon, Haffert, Herscovici-Schiller, Hours, Juanola-Parramon, Kleisioti, König, van Kooten, Krasteva, Laginja, Landman, Leboulleux, Mouillet, N’Diaye, Por, Pueyo, \& Snik}]{Vaughn_2023}
Vaughan, S.~R., Gebhard, T.~D., Bott, K., {et~al.} 2023, Monthly Notices of the Royal Astronomical Society, 524, 5477, \dodoi{10.1093/mnras/stad2127}

\bibitem[{Virtanen {et~al.}(2020)Virtanen, Gommers, Oliphant, Haberland, Reddy, Cournapeau, Burovski, Peterson, Weckesser, Bright, {van der Walt}, Brett, Wilson, Millman, Mayorov, Nelson, Jones, Kern, Larson, Carey, Polat, Feng, Moore, {VanderPlas}, Laxalde, Perktold, Cimrman, Henriksen, Quintero, Harris, Archibald, Ribeiro, Pedregosa, {van Mulbregt}, \& {SciPy 1.0 Contributors}}]{2020SciPy-NMeth}
Virtanen, P., Gommers, R., Oliphant, T.~E., {et~al.} 2020, Nature Methods, 17, 261, \dodoi{10.1038/s41592-019-0686-2}

\bibitem[{Waite {et~al.}(2007)Waite, Young, Cravens, Coates, Crary, Magee, \& Westlake}]{Waite_2007}
Waite, J.~H., Young, D.~T., Cravens, T.~E., {et~al.} 2007, Science, 316, 870, \dodoi{10.1126/science.1139727}

\bibitem[{Wang {et~al.}(2017)Wang, Mawet, Ruane, Hu, \& Benneke}]{Wang_2017}
Wang, J., Mawet, D., Ruane, G., Hu, R., \& Benneke, B. 2017, The Astronomical Journal, 153, 183, \dodoi{10.3847/1538-3881/aa6474}

\bibitem[{West \& Smith(1991)}]{WEST1991330}
West, R.~A., \& Smith, P.~H. 1991, Icarus, 90, 330, \dodoi{https://doi.org/10.1016/0019-1035(91)90113-8}

\bibitem[{Williams \& Gaidos(2008)}]{Williams_2008}
Williams, D.~M., \& Gaidos, E. 2008, Icarus, 195, 927, \dodoi{https://doi.org/10.1016/j.icarus.2008.01.002}

\bibitem[{Xie \& Ji(2002)}]{xie_2002}
Xie, Y., \& Ji, Q. 2002, in 2002 International Conference on Pattern Recognition, Vol.~2, IEEE, 957--960

\bibitem[{Yahn {et~al.}(2025)Yahn, Trent, Duncan, Seignovert, Santerre, \& Nixon}]{Yahn_2025}
Yahn, Z., Trent, D.~M., Duncan, E., {et~al.} 2025, Journal of Geophysical Research: Machine Learning and Computation, 2, \dodoi{10.1029/2024jh000366}

\bibitem[{{Zebker} {et~al.}(2009){Zebker}, {Stiles}, {Hensley}, {Lorenz}, {Kirk}, \& {Lunine}}]{Zebker_2009}
{Zebker}, H.~A., {Stiles}, B., {Hensley}, S., {et~al.} 2009, Science, 324, 921, \dodoi{10.1126/science.1168905}

\bibitem[{{Zhang} {et~al.}(2022){Zhang}, {Knutson}, {Wang}, {Dai}, \& {Barrag{\'a}n}}]{Zhang_2022}
{Zhang}, M., {Knutson}, H.~A., {Wang}, L., {Dai}, F., \& {Barrag{\'a}n}, O. 2022, \aj, 163, 67, \dodoi{10.3847/1538-3881/ac3fa7}

\bibitem[{Zugger {et~al.}(2010)Zugger, Kasting, Williams, Kane, \& Philbrick}]{Zugger_2010}
Zugger, M.~E., Kasting, J.~F., Williams, D.~M., Kane, T.~J., \& Philbrick, C.~R. 2010, The Astrophysical Journal, 723, 1168, \dodoi{10.1088/0004-637X/723/2/1168}

\end{thebibliography}
\bibliographystyle{aasjournal}

\end{document}